\documentclass[]{svjour3}
\usepackage{mydef}

\usepackage{tikz}
\usetikzlibrary{calc,decorations.markings}
\usetikzlibrary{arrows,patterns,positioning}

\def\tmm{\tilde{\mu}_1^<}
\def\taups{\tau^\prime-s}
\def\bari{\bar{i}}

\begin{document}


\title{{\itshape Structural default model with mutual obligations}}

\author{Andrey Itkin \and Alexander Lipton
\thanks{The views expressed in this paper represent those of the authors and not necessarily those of Bank of America}}

\institute{A. Itkin \at
Bank of America Merrill Lynch, New York NY, USA \\
\& New York University, School of Engineering,\\
\email{aitkin@nyu.edu} 
\and
A. Lipton \at
Bank of America Merrill Lynch, New York NY, USA \\
\& Oxford-Man Institute of Quantitative Finance, University of Oxford, UK, \\
\email{alex.lipton@baml.com}
}



\date{\today}

\maketitle

\begin{abstract}
This paper considers mutual obligations in the interconnected bank system and
analyzes their influence on joint and marginal survival probabilities as well as CDS and FTD prices
for the individual banks. To make the role of mutual obligations more transparent, a simple
structural default model with banks' assets driven by correlated multidimensional Brownian motion
with drift is considered. This model enables a closed form representation for
many quantities of interest, at least in a 2D case, to be obtained, and moreover, model calibration
is provided. Finally, we demonstrate that mutual obligations have to be taken into account in order to
get correct values for model parameters.



\end{abstract}

\section{Introduction}
Structural default framework is widely used for assessing credit risk of a corporate debt. It was introduced in a simple form in the seminal work \cite{m74}, and was further extended in various
papers, see a survey in \cite{LiptonSepp2011} and references therein. In contrast to reduced-form
models (see, e.g., \cite{Bielecki2004}) structural default models suffer from the curse of dimensionality when the number of counterparties grows; however, these models provide a more detailed and financially meaningful description of the default event for a typical firm.

Inspired by \cite{BOE2011}, in \cite{ItkinLipton2014} we extended the structural
framework by taking into account the fact that banks have mutual liabilities among themselves. Accounting for this effect is very important in order to accurately analyze credit worthiness of individual
banks and the banking network as a whole. For instance, large mutual liabilities imply that adverse shock to a bank is rapidly transmitted to the entire system, with severe implications for its stability (\cite{DavidLehar2014}). The authors of \cite{DavidLehar2014} indicate that renegotiations between highly interconnected banks facilitate mutual private sector bailouts to lower the need for government bailouts. The relative size of mutual liabilities compared to total liabilities is fairly substantial. For instance, the relative fraction of interbank loans is 12\% in the EU, 8.5\% for Canada (\cite{DavidLehar2014}), and 4.5\% for US (as per Economic Research website of the Federal Reserve Bank of St. Louis).

An extended Merton model with mutual liabilities and continuous default monitoring can be built by
combining correlated Merton balance sheet models, calibrated by using observed bank equity returns,
and analyzing potential clearing of interbank liabilities in the spirit of \cite{EN2001}. In \cite
{ItkinLipton2014} we assumed that banks' assets are driven by correlated L\'{e}vy processes with idiosyncratic and common components and developed a novel pseudo partial differential
equation computational method in order to make the problem of computing joint and marginal survival
probabilities tractable. The effect of mutual liabilities was discussed, and numerical examples were
given to illustrate its importance.

Obviously, the knowledge of joint and marginal survival probabilities is important for successful
calibration of the model to Credit Default Swap (CDS) spreads and first-to-default (FTD) instruments. Since the general case is fairly complicated, here we restrict ourselves to the case when banks' assets are driven by correlated Brownian Motions with drifts. Then in the 2D case we obtain explicit expressions for
several quantities of interest including joint and marginal survival probabilities as well as CDS
and FTD prices. Despite the fact that the model under consideration does not incorporate jumps, it is still beneficial as it enables an analytical assessment and provides a natural link to the analytical framework considered in \cite{LiptonSepp2011, LiptonSavescu2014}.

The rest of the paper is organized as follows. In Section~\ref{Model} we propose a model for the general
case of $N$ banks. Sections~\ref{ge},\ref{Surv} present the governing equations and Green's
function approach to the solution of these equations for joint and marginal survival probabilities
for two banks with mutual obligations. In Sections~\ref{sCDS},\ref{sFTD} the prices of CDS and FTD contracts are calculated, and results of our numerical experiments are presented. We also validate our results by comparing analytical solutions with numerical solutions obtained by using a finite
difference algorithm described in \cite{ItkinLipton2014}. Section~\ref{Calib} discusses calibration
of the model presents some numerical results. Our conclusions are presented in Section~\ref{Concl}.

\section{Model} \label{Model}
Consider a set of $N$ banks with external assets and liabilities $A_{i}$, $L_{i}$, $i=1,...,N$, and interbank assets and liabilities $L_{ji}$, $j=1,...,N$, respectively.
In other words, $L_{ij}$ is the amount the $i$-th bank owes from the $j$-th bank, etc.
Thus, total assets, liabilities and capital of the $i$-th bank have the form
\begin{align*} 
\widetilde{A}_{i} &= A_{i}+\sum_{j\neq i}L_{ji},  \qquad 
\widetilde{L}_{i} = L_{i}+\sum_{j\neq i}L_{ij},  \qquad
C_{i} = \widetilde{A}_{i}-\widetilde{L}_{i} = A_{i} - \lambda_{i}^{=}, \nonumber
\end{align*}
\noindent where
\begin{equation} \label{Eq2}
\lambda_{i}^{=} = L_{i} + \sum_{j\neq i}\left( L_{ji} - L_{ij}\right).
\end{equation}

For simplicity, we assume that the corresponding dynamics is governed by the
SDEs of the form
\begin{align} \label{govern}
dA_{i, t}  &= r A_{i,t} dt + \sigma_i A_{i,t} d W_{i,t}, \\
dL_{i,t}   &= r L_{i,t} dt, \qquad dL_{ij,t}  = r L_{ij,t} dt, \nonumber
\end{align}
\noindent subject to the initial conditions $A_{i,0} = A_i(0), \ L_{i,0} = L_i(0), \ L_{ij,0} = L_{ij}(0)$, so that $L_{i,t}, L_{ij,t}$ are deterministic functions of the time $t$\footnote{Accordingly, further we will denote them as $L_i(t), \ L_{ij}(t)$.}. In \eqref{govern} $r$ is the risk-free rate, $\sigma_i$ is the volatility of the $i$-th asset (which is assumed to be constant), and $W_{i,t}$ is the corresponding Brownian Motion. Elements of the corresponding correlation matrix are denoted by $\rho_{ij}$. The above assumptions can be generalized in a variety of ways, which will
be discussed elsewhere. 

We assume that all the liabilities (both external and interbank) are settled at a certain maturity $T > 0$. Thus at $t=T$, the $i$-th bank defaults if $\widetilde{A}_{i,T} < \widetilde{L}_i(T)$, or, equivalently, if $A_{i,T} < \lambda_i^=(T)$. Below we denote the default time of the $i$th bank by $\tau_i$.

\paragraph{The $k$-th bank defaults at $\tau_k < T$.}
We describe defaults at intermediate times $0 < \tau_i < T$ in the spirit of \cite{BC1976}
by assuming that the $i$-th bank defaults at time $\tau_i$ provided that
\begin{align} \label{defBound}
A_i(\tau_i) & \le \lambda_i^< (\tau_i), \\
\lambda_i^< (\tau_i) &= R_i \Big[ L_i(\tau_i) + \sum_{j \ne i} L_{ij}(\tau_i) \Big]
- \sum_{j \ne i} L_{ji}(\tau_i), \nonumber
\end{align}
\noindent where $0 \le R_i \le 1$ is the recovery rate, which is assumed to be constant up to the time $t = T$. We need to emphasize that according to these settings, the default boundary is discontinuous at $t=T$, because $R_i$ experiences a jump at this point from its value $R_i$ at $t  < T$ to 1 at $t=T$ (and so $\lambda_i^<$ transforms to $\lambda_i^=$).

These default boundaries are valid if no other bank defaults till $t=T$. Now assume the opposite, i.e. that
the $k$-th bank is the first to default at time $\tau_k < T$, and so we are left with a
set of $N-1$ surviving banks. At time $\tau_k$ the assets and liabilities of the $i$-th bank, $i \ne k$, have the form
\begin{align*}
\widetilde{A}_i(\tau_k) &= A_i(\tau_k) + \sum_{j\ne i,j\ne k}L_{ji}(\tau_k) + R_k L_{ki}(\tau_k), \\
\widetilde{L}_{i}(\tau_k) &= L_i(\tau_k) + \sum_{j\ne i,j\ne k} L_{ij}(\tau_k) + L_{ik}(\tau_k). \nonumber
\end{align*}

We assume that for surviving banks mutual liabilities stay the same, while their external liabilities jump according to the rule
\begin{equation*}
L_i (\tau_k) \rightarrow \bar{L}_i(\tau_k) \equiv L_i(\tau_k) + L_{ii}(\tau_k) - R_k L_{ki}(\tau_k).
\end{equation*}

Surviving banks' capital naturally takes a hit%
\begin{equation*}
C_i(\tau_k) \rightarrow \bar{C}_i(\tau_k) \equiv C_i(\tau_k) - (1-R_k) L_{ki}(\tau_k).
\end{equation*}

Thus, each default reduces the set of surviving banks and modifies the corresponding default
boundaries as
\begin{align} \label{defBoundMutual}
\tilde{\lambda}_{ik}(t) &=
\begin{cases}
\lambda_{ik}^<(t), & t < T, \cr
\lambda_{ik}^=(T), & t = T, \cr
\end{cases}
\quad i \ne k \\
\lambda _{ik}^<(t) &= R_i \left[ L_i(t) + L_{ik}(t) - R_{k}L_{ki}(t) \right] 
+ \sum_{j\ne i, j\ne k} \left[R_i L_{ij}(t) - L_{ji}(t) \right], \nonumber \\
\lambda _{ik}^=(T) &= L_i(t) + L_{ik}(t) - R_{k}L_{ki}(t)
+ \sum_{j\ne i, j\ne k} \left[L_{ij}(t) - L_{ji}(t) \right], \nonumber
\end{align}

As an example, consider $N=2$. Then for the remaining bank $\bar{k} \equiv 3-k$, the default boundary is given by (\cite{ItkinLipton2014})\footnote{In this case we omit the second index of $\lambda_{ik}$ as the defaulted bank is determined uniquely.}
\begin{equation} \label{defBoundMutual}
\tilde{\lambda}_{\bar{k}}(t)=
\begin{cases}
R_{\bar{k}}\left( L_{\bar{k}}+L_{\bar{k}k}-R_{k}L_{k\bar{k}}\right), & \tau_k \le t < T, \cr
L_{\bar{k}}+L_{\bar{k}k}-R_{k}L_{k\bar{k}}, & \tau_k < t = T.
\end{cases}
\end{equation}

It is clear that
\begin{equation*} \label{delLambda}
\Delta \lambda_{\bar{k}} =\tilde{\lambda}_{\bar{k}}-\lambda _{\bar{k}}=
\begin{cases}
\left( 1-R_{\bar{k}}R_{k}\right) L_{k\bar{k}} \equiv \Delta \lambda^<_{\bar{k}}, & \tau_k \le t < T, \cr
\left( 1-R_{k}\right) L_{k\bar{k}} \equiv \Delta \lambda^=_{\bar{k}}, & t = T,
\end{cases}
\end{equation*}
\noindent and $\Delta \lambda_{\bar{k}} \ge 0$.

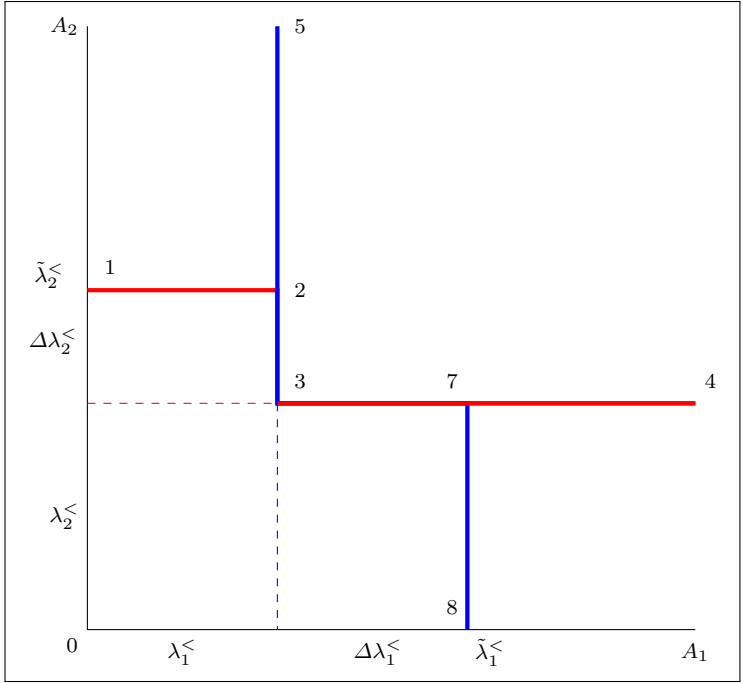
\begin{figure}[!h]
\begin{center}
\fbox{
\begin{tikzpicture}
\def\sizeG{8.};
\def\lA{2.5};
\def\lB{3.};
\def\lAB{2.};
\def\lBA{1.5};
\draw (0,0) -- (\sizeG,0)
      (0,0) -- (0,\sizeG);
\draw[red, ultra thick] (0,\lB + \lBA) -- (\lA,\lB + \lBA) -- (\lA,\lB) -- (\sizeG,\lB);
\draw[blue, ultra thick] (\lA,\sizeG) -- (\lA,\lB) -- (\lB + \lAB,\lB) -- (\lB + \lAB,0);
\draw[red, ultra thick] (\lA,\lB) -- (\sizeG,\lB);
\draw[red, dashed] (0,\lB) -- (\lA,\lB);
\draw[blue, dashed] (\lA,\lB) -- (\lA,0);
\node at (\sizeG,-0.3){$A_1$};
\node at (-0.3,\sizeG){$A_2$};
\node at (-0.2,-0.2){$0$};
\node at (0.5*\lA,-0.3){$\lambda_1^<$};
\node at (-0.3,0.5*\lB){$\lambda_2^<$};
\node at (0.85*(\lA+\lAB,-0.3){$\Delta \lambda_1^<$};
\node at (-0.45,0.85*(\lB+\lBA){$\Delta \lambda_2^<$};
\node at (\lA + 0.3,\lB+0.3) {$3$};
\node at (\lA + \lAB + 0.3,\lB+0.3) {$7$};
\node at (\sizeG + 0.2,\lB+0.3) {$4$};
\node at (0.3,\lB + \lBA+0.3) {$1$};
\node at (-0.5,\lB + \lBA+0.2) {$\tilde \lambda_2^<$};
\node at (\lA + 0.3,\lB + \lBA) {$2$};
\node at (\lA + \lAB + 0.3,0.3) {$8$};
\node at (\lA + \lAB + 0.8,-0.3) {$\tilde \lambda_1^<$};
\node at (\lA + 0.3,\sizeG) {$5$};
\end{tikzpicture}
}
\end{center}
\caption{Default boundaries of two banks with and without mutual liabilities at $t < T$.}
\label{Fig1}
\end{figure}

To make these definitions more transparent the computational domain is represented in Fig.~\ref{Fig1}. Here, if there are no defaults, we have a rectangular computational domain which lies above the piece-wise constant line $5-3-4$. If the bank 2 defaults, this domain transforms to that which lies to the right of the line $5-3-7-8$. If the bank 1 defaults, the domain transforms to that which lies above the line $1-2-3-4$.

\paragraph{The $i$-th bank defaults at $\tau_i = T$.} 
In this case the definition of $\tilde{\lambda}_i$ in \eqref{defBoundMutual} 
should be changed. Indeed, if assets of the $i$-th bank breach below its liabilities at some time before maturity, the bank has some period of time to recover, unless it breaches below the level $\tilde \lambda_i^<$. At this level the bank's counterparties don't believe anymore in its ability to recover, and it defaults. Obviously, at $t=T$ the bank doesn't have time to recover. Therefore, at the most it can pay to its obligors the current amount of money in hands, i.e. the total value of the bank assets\footnote{
We consider an idealistic situation when all bank's assets upon default can be immediately converted to cash with no delay and further losses.} which is a fraction $\gamma_i, \ 0 < \gamma_i \le 1$, of its liabilities. Accordingly, in the spirit of \cite{EN2001}, to determine default boundaries we need to find the vector $\boldsymbol{\gamma} = \{\gamma_i\}, \ 0 \le \gamma_i \le 1, \ i \in [1,N]$ which solves the following piece-wise linear problem in the unit cube:
\begin{equation} \label{gamProblem}
\min\left\{A_i(T) + \sum_{j\ne i} \gamma_j L_{ji}(T), \ L_i(T) + \sum_{i \ne j} L_{ij}(T) \right\} 
 = \gamma_i \left( L_{i}(T) +\sum_{j\ne i}L_{ij}(T) \right).
\end{equation}

Introducing new non-dimensional variables $a_i = A_i(T) / \tilde{L}_{i}(T), \ l_{ji} = L_{ji}(T) / \tilde{L}_{i}(T)$
the problem given in \eqref{gamProblem} can be re-written in the form
\begin{equation} \label{gamPr2}
\min \left\{ a_{i} + \sum_{j\ne i} \gamma_j l_{ji}, \ 1 \right\} = \gamma_i.
\end{equation}

It is clear that $\gamma_i = 1$ (so that the $i$-th bank survives) if $a_{i} + \sum_{j\ne i} \gamma_j l_{ji} \ge 1$. And $\gamma _i < 1$ otherwise, so that the $i$-th bank defaults. This description suggests that defaults in the interlinked set of banks can happen outright, when
\begin{equation*}
A_{i}\left( T\right) <\lambda _{i}^{=}\left( T\right),
\end{equation*}%
\noindent and through contagion, when
\begin{equation*}
L_i (T) +\sum_{j\neq i}\left[ L_{ij}\left( T\right) -\gamma _{j}L_{ji}\left( T\right) \right] > A_{i}\left( T\right) \geq \lambda _{i}^{=}\left( T\right).
\end{equation*}
\eqref{gamPr2} can be uniquely solved. A brief discussion is given in Appendix~\ref{App0}.

Accordingly, in this case we change the definition of $\lambda^=_i(T)$ at $\tau_i = T$ to 
\begin{equation} \label{defBoundMutualT}
\tilde{\lambda}_{i,T} = L_{i}(T) + \sum_{i\ne j} L_{ij}(T) - \sum_{j\ne i} \tilde{R}_{j,T}(\boldsymbol{\gamma})L_{ji}(T)
\end{equation}
\noindent where
\begin{equation} \label{tildR}
\tilde{R}_{j,T}(\boldsymbol{\gamma})= \min \left[ 1, \ \dfrac{A_{j}(T) + \sum_{i \ne j} \gamma_i L_{ij}(T)}{L_{j}(T) + \sum_{i \ne j} \gamma_i L_{ji}(T)} \right], \qquad \boldsymbol{\gamma} = [\gamma_1,...,\gamma_N].
\end{equation}

\begin{figure}[!h]
\begin{center}
\fbox{

\begin{tikzpicture}
\def\sizeG{8.};
\def\lA{2.5};
\def\lB{3.};
\def\lAB{2.};
\def\lBA{1.5};
\def\sh{0.03};
\def\sH{0.03};
\def\dis{1.5};

\draw (0,0) -- (\sizeG,0)
      (0,0) -- (0,\sizeG);
\draw[red, ultra thick] (0,\lB + \lBA*2.5) -- (\lA,\lB) -- (\sizeG,\lB);
\draw[red, dashed] (0,\lB) -- (\lA + \lAB + 0.8,\lB);
\draw[red, dashed] (0,\lB-1) -- (\sizeG,\lB - 1);
\draw[red,thick] (\lA-\dis,\lB-1) -- (\sizeG,\lB - 1);
\draw[red, dashed] (0,\lB-1) -- (\sizeG,\lB - 1);
\draw[red, dashed] (\lA, \lB + \lBA+0.8) -- (0, \lB + \lBA+0.8);
\draw[red, dashed] (0,\lB - 0.5) -- (\sizeG,\lB - 0.5);

\draw[blue, ultra thick] (\lA,\sizeG) -- (\lA,\lB) -- (\lA + \lAB*2.,0);
\draw[blue, dashed] (\lA,\lB+\lBA) -- (\lA,0);
\draw[blue, dashed] (\lA + \lAB-0.6,\lB) -- (\lA + \lAB-0.6,0.0);
\draw[blue, dashed] (\lA-\dis, 0) -- (\lA-\dis,\sizeG);
\draw[blue,thick] (\lA-\dis, \lB-1) -- (\lA-\dis,\sizeG);
\draw[blue, dashed] (\lA - 0.5*\dis, 0) -- (\lA - 0.5*\dis, \sizeG);

\node at (\sizeG,-0.3){$A_1$};
\node at (-0.3,\sizeG){$A_2$};
\node at (-0.2,-0.2){$0$};
\node at (\lA,-0.3){$\lambda_1^=$};
\node at (-0.45,\lB){$\lambda_2^=$};
\node at (\lA+0.3,\lB+0.3){$6$};
\node at (\lA+\lAB - 0.6,\lB+0.3){$7$};

\node at (-0.45,(\lB-1){$\lambda_2^<$};
\node at (-0.45,(\lB-0.5){$\tilde \lambda_2^<$};
\node at (\lA-\dis,-0.3){$\lambda_1^<$};
\node at (\lA-0.5*\dis,-0.3){$\tilde \lambda_1^<$};

\node at (\lA-\dis-0.2,(\lB-1-0.2){$3$};
\node at (\sizeG - 0.2,\lB+0.3) {$4$};
\node at (\lA-\dis+0.3,\lB+\lBA+1) {$2$};
\node at (\lA + \lAB-0.5, \lB-0.8) {$9$};
\node at (\lA + 0.3,\sizeG-0.3) {$5$};
\node at (\lB + \lAB+1.6,0.3) {$8$};
\node at (0.2,\lB + \lBA*2.5) {$1$};
\node at (\lA + \lAB-0.6,-0.3) {$\tilde \lambda_1^=\big|_{t^d < T}$};
\node at (-0.7,\lB + \lBA+0.8) {$\tilde \lambda_2^=\big|_{t^d < T}$};

\node at (\lA + 1.5*\lAB, \lB + 1.5*\lBA) {$\mathbf{D_{12}}$};
\node at (\lA + 2*\lAB, \lB - 0.5) {$\mathbf{D_{1}}$};
\node at (\lA - 0.5, \lB + 2*\lBA) {$\mathbf{D_{2}}$};
\node at (\lA-0.5 , \lB - 0.5) {$\hat{\mathbf D}$};

\draw[pattern=dots] (\lA-\dis,\lB-1) rectangle (\sizeG,\sizeG);
\end{tikzpicture}
}
\end{center}
\caption{Default boundaries of two banks with and without mutual liabilities at $t = T$. The dot pattern marks the whole computational domain $\mathfrak{D}$.}
\label{Fig2}
\end{figure}
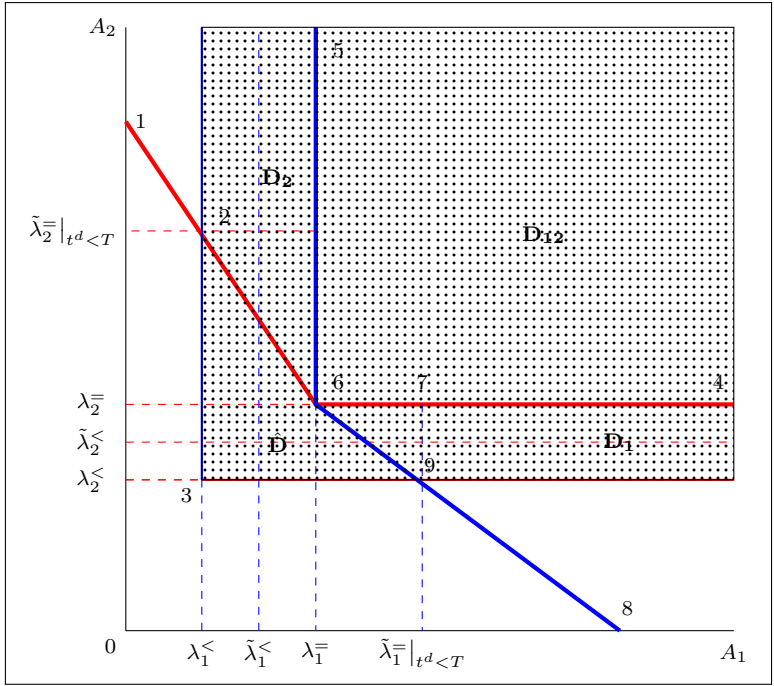

It follows that the default boundary $\tilde\lambda_{i,T}$ piece-wise linearly depends on all $A_j(T), \ j \in[1,N], \ j \ne i$. In particular, let $N=2$ and $\tau_2 = T$, hence when $A_2(T) = 0$ we have from \eqref{defBoundMutualT}
\[ \tilde\lambda_{1,T} = \dfrac{\bigtriangleup}{L_2 + L_{21}}, \quad  \bigtriangleup = L_1 L_2 + L_{12} L_2 + L_1 L_{21}. \]
Therefore,
\[ \tilde\lambda_1^=\Big|_{\tau_2 < T}  - \tilde\lambda_{1,T} = L_{21}(\tilde R_{2,T}(1) - R_2) = L_{21}\left( \dfrac{L_{12}}{L_2 + L_{21}} - R_2 \right). \]

This behavior is illustrated in Fig.~\ref{Fig2}. Since $\tau_j = T$, and, thus, $R_j = 1$, from \eqref{defBoundMutual} we have $\tilde \lambda_i^= = \lambda_i^=$, or $\Delta \lambda_i = 0$.
Therefore, when $A_2(T)$ grows from 0 to $\lambda_2^=$, the default boundary $\lambda_{1,T}$ moves from $\dfrac{\bigtriangleup}{L_2 + L_{21}}$ to $\lambda_1^=$ along the line 8-9-6.

At point 6 the default boundary $\tilde\lambda_{1,T}$ transforms to $\tilde\lambda_1^= = \lambda_1^=$, and further doesn't depend on $A_2(T)$ when the latter increases. This occurs at the point $A^=_2(T) = \lambda_2^=$. Thus, the whole default boundary of the first bank in Fig.\ref{Fig2} can be seen as a line passing through the points 8-9-6-5. Similarly, for the second bank the default boundary in Fig.~\ref{Fig2} can be seen as a line passing through the points 1-2-6-4.

Also in Fig.~\ref{Fig2} $\mathbf{D}_{12}$ is the domain where both banks don't default, $\mathbf{D}_1$ - where the second bank defaults while the first one does not, $\mathbf{D}_2$ - where the first bank defaults while the first one does not, $\mathfrak{D}$ is the whole computational domain marked by the dot pattern, and in the domain ${\mathbf D} = \mathfrak{D}\backslash\mathbf{D}_{12}$ both banks default.

As always, it is useful to describe the evolution of the set of banks under consideration in terms of non-dimensional variables. To this end, we introduce the average volatility $\omega \equiv \left(\prod_{i-1}^N \sigma_i\right)^{1/N}$, and define
\[ \bar{t} = \omega^2 t, \quad \bar{T} = \omega^2 T, \quad X_{i, \bar{t}} = \dfrac{\omega}{\sigma_i}\ln
\left(\dfrac{A_{i, \bar{t}}}{\lambda_i^<(\bar{t})} \right), \quad \xi_i = -\dfrac{1}{2}\dfrac{\sigma_i}{\omega}. \]
The corresponding dynamics of $\bar{X}_{i, \bar{t}}$ is governed by the SDE:
\begin{equation} \label{dynX}
d\bar{X}_{i,\bar{t}}= \xi_i d \bar{t} + d W_{i,\bar{t}},
\end{equation}
\noindent while the default conditions now transform to $X_{i}\leq \mu _{i}$, with $\mu_{i}$ defined as
\begin{equation} \label{mu}
\mu_{i}\left( \bar{t}\right) =
\begin{cases}
\mu _{i}^{<} \equiv 0,  &  \bar{t} < \bar{T}, \cr
\mu_{i}^{=} \equiv \dfrac{\omega}{\sigma_i}
\ln \left(\dfrac{\lambda_i^=(\bar{t})}{\lambda_i^<(\bar{t})}\right), & \bar{t} = \bar{T}.
\end{cases}
\end{equation}
\noindent By definition, $\mu_{i}^{=} > 0$.

Below we omit bars for the sake of simplicity.

If the $j$-th bank defaults at $\tau_j < T$, then for the $i$-th bank the default boundary is
given by
\begin{equation} \label{Eq10}
\tilde{\mu}_{ij}(t) =
\begin{cases}
\tilde{\mu}_{ij}^< \equiv \dfrac{\omega}{\sigma _{i}}\ln \left(\dfrac{\tilde\lambda_{ij}^<(t)}{\lambda_{ij}^<(t)}\right),
& t<T, \cr
\tilde{\mu}_{ij}^= \equiv \dfrac{\omega}{\sigma _{i}}\ln \left(\dfrac{\tilde\lambda_{ij}^=(t)}{\lambda_{ij}^<(t)}\right), & t=T.
\end{cases}
\end{equation}
Note, that according to \eqref{govern} $\tilde{\mu}_{ij}^<$ doesn't depend on $t$.

It can be seen that the boundary condition in \eqref{Eq10} at $t=T$ doesn't match to the terminal condition which, according to \eqref{defBoundMutualT}, reads
\begin{equation} \label{muTilT}
\tilde{\mu}_{ij,T} = \dfrac{\omega}{\sigma _{i}}\ln \left(\dfrac{\tilde\lambda_{ij,T}(T)}{\lambda_{ij}^<(T)}\right) \ne \tilde{\mu}_{ij}^=(T).
\end{equation}

Mathematically, this means that our problem belongs to the class of problems with a boundary (transition) layer at $t=T$. Financially, the behavior of the solution in this layer is determined by the detailed specification of the contract. For instance, if the bank is close to maturity, say a day before, the recovery rate could be defined to smoothly transit from $R_i$ to 1 within this last day. Or, some other conditions specific to the contract in question could be issued. However, we don't consider these details, assuming that the boundary layer is thin, and, therefore, any perturbation of the solution due to the existence of this layer, is dumped out pretty fast when moving away from this layer. In other words, as we ignore a detailed consideration of the boundary layer, our solution experiences a jump at $t=T$. Therefore, after the closed form solution is found we will compare it with the numerical solution of this problem to reveal sensitivity of the former to the value of the described effect.

Below we provide all the results just for two-dimensional case $N=2$ while the multi-dimensional case will be presented elsewhere. Accordingly, in the definition of $\tilde{\mu}_{ij}^=, \ \tilde{\mu}_{ij}^<$ for easiness of reading we will omit the second index as in this case it doesn't bring any confusion. 

\section{Governing equations} \label{ge}
Based on the analysis presented in the previous section, the joint survival probability $Q\left( t,X_{1},X_{2}\right)$ of two assets $X_1,X_2$ is defined in the domain $\Omega(t,X_1,X_2): [0,T] \times [0,\infty] \times [0,\infty]$\footnote{The space sub-domain of $\Omega$ corresponds to the dotted area in Fig.~\ref{Fig2}.}. It solves the following terminal boundary value problem (\cite{LiptonSepp2011})
\begin{align} \label{Eq11}
Q_{t}\left( t,X_{1},X_{2}\right) +\mathcal{L}Q\left( t,X_{1},X_{2}\right) &= 0, \\
Q\left( T,X_{1},X_{2}\right) = \mathbf{1}_{X\in \mathbf{D}_{12}}, \quad
Q\left( t,0,X_{2}\right) &= 0, \quad Q\left( t,X_{1},0\right) = 0,  \nonumber
\end{align}
\noindent where
\begin{align*}
\mathcal{L}Q &= \Delta_\rho Q + \xi \cdot \nabla Q,   \noindent \\
\Delta_\rho &\equiv \dfrac{1}{2} \sop{}{X_1} + \rho \msop{}{X_1}{X_2} + \dfrac{1}{2} \sop{}{X_2}, \qquad
\xi = \left(\xi_1, \xi_2\right)^T, \noindent
\end{align*}
\noindent $\mathbf{1}_x$ is the Heaviside step function defined with the half-maximum convention
\footnote{Since the detailed consideration of the transition layer at $t=T$ is omitted, this condition allows getting the correct value of $\chi_1(T,\tilde\mu_1^=)$, see \eqref{chi1}.},
and the area $\mathbf{D}_{12}$ is defined in Fig.~\ref{Fig2}. We emphasize that the domain $\mathbf{D}_i$ in $\vec{X}$ variables has a curvilinear boundary which depends on the value of $A_{\bari}(T)$. Indeed, based on the definitions in \eqref{defBound}, \eqref{defBoundMutualT}, \eqref{tildR} and \eqref{muTilT}, one can find, e.g., for $i=1$
\begin{equation*}
\tilde \mu_{1,T} = \dfrac{\omega}{\sigma_1} \ln \left[\dfrac{L_1 + L_{12} - \tilde{R}_{2,T}(1) L_{21}}{L_1 + L_{12} - L_{21}} \right] = \dfrac{\omega}{\sigma_1} \ln \left[1 + \dfrac{L_{21}(1 - \tilde{R}_{2,T}(1))}{L_1 + L_{12} - L_{21}}\right].
\end{equation*}
%

Next we define the corresponding marginal survival probabilities $q_{i}\left(t,X_{1},X_{2}\right)$, $i=1,2$, which are functions of \emph{both} $X_{1}$ and $X_{2}$, also in the domain $\Omega(t,X_1,X_2)$. For brevity we provide all definitions and formulae for the first bank ( $i=1$) while for the second one it could be done by analogy. So $q_1(t,X_{1},X_{2})$ solves the following terminal boundary value problem
\begin{align} \label{Eq13}
q_{1,t}\left( t,X_{1},X_{2}\right) &+ \mathcal{L}q_{1}\left(t,X_{1},X_{2}\right) = 0,  \\
q_{1}\left( T,X_{1},X_{2}\right) &= \mathbf{1}_{\mathbf{X} \in [\mathbf{D}_{12} \cup \mathbf{D}_1]},   \nonumber \\
q_{1}\left( t,0,X_2\right) = 0, \quad
q_{1}\left( t,X_1,0\right) &\equiv \Xi(t,X_1) =
\begin{cases}
\chi_{1,0}( t,X_1), & X_1 \geq \tilde \mu_1^{<}, \cr
0, & 0 \le X_1 < \tilde \mu_1^<,
\end{cases}, \nonumber \\
q_{1}\left( t,X_1,X_2\uparrow\infty\right) &= \chi_{1,\infty}(t,X_1),
\quad q_{1}\left( t,X_1\uparrow\infty, X_2\right) = 1.
 \nonumber
\end{align}

In \eqref{Eq13} the domain $\mathbf{D}_1$ is defined in Fig.~\ref{Fig2}.
Function $\chi_{1,0}( t,X_1)$ is the 1D survival probability, which solves the following terminal boundary value problem
\begin{align} \label{Eq14}
\partial_t \chi_{1,0}( t,X_1) + \mathcal{L}_{1} \chi_{1,0}(t,X_1) &= 0, \\
\chi_{1,0}( T,X_1) = \mathbf{1}_{X_1 > \tilde{\mu}_1^=},  \quad
\chi_{1,0}( t,\tilde\mu_{1}^{<}) &= 0, \nonumber
\end{align}
\noindent where
\begin{equation*} \label{Eq15}
\mathcal{L}_i = \frac{1}{2} \sop{}{X_i} + \xi_i\fp{}{X_i}.
\end{equation*}

Accordingly, function $\chi_{1,\infty}( t,X_1)$ is the 1D survival probability, which solves the following terminal boundary value problem
\begin{align} \label{Eq14a}
\partial_t \chi_{1,\infty}( t,X_1) + \mathcal{L}_{1} \chi_{1,\infty}(t,X_1) &= 0, \\
\chi_{1,\infty}( T,X_1) = \mathbf{1}_{X_1 > \mu_1^=},  \quad
\chi_{1,\infty}( t,0) &= 0, \nonumber
\end{align}

\section{Survival probabilities} \label{Surv}
We solve \eqref{Eq11} and \eqref{Eq13} by introducing the Green's function $G(t,X_1,X_2 | t', X_1^\prime, \\ X_2^\prime)$, where $X_1', X_2'$ are the initial values of $X_1,X_2$ at $t=t'$. Below, where it is not confusing, for brevity we will also use the notation $G(t-t',X_1,X_2)$, thus explicitly exploiting the fact that for our problem the Green's function depends only on $t-t'$, and omitting the second pair of arguments. The Green's function solves the following initial boundary value problem
\begin{align} \label{Eq16}
G_t(t-t',X_{1},X_{2}) &- \mathcal{L}^\dag G(t-t',X_1,X_2) = 0,   \\
G(0,X_1,X_2) &= \delta (X_1-X_1^\prime) \delta( X_2-X_2^\prime), \nonumber \\
G(t-t',0,X_2) &= 0, \quad G( t-t',X_1,0) = 0, \nonumber
\end{align}
\noindent where $\mathcal{L}^{\dag } = \Delta _{\rho } - \xi \cdot \nabla$. A simple calculation yields
\begin{equation*}
\left( Q G\right) _{t}+\mathcal{L} G Q - Q \mathcal{L}^{\dag }G = 0,
\end{equation*}
\noindent  or, explicitly,
\begin{equation*}
\left( QG\right) _{t}+\nabla \cdot \left(
\begin{array}{c}
\dfrac{1}{2}\left( Q_{X_{1}} G - Q G_{X_{1}}\right) -\rho Q G_{X_{2}} + \xi_{1} Q G \\
\dfrac{1}{2}\left( Q_{X_{2}} G - Q G_{X_{2}}\right) +\rho Q_{X_{1}} G + \xi _{2}Q G
\end{array}
\right) =0.
\end{equation*}

The Green's theorem (\cite{Kythe2011}) yields
\begin{align} \label{Eq20}
Q \left(t',X_1^\prime,X_2^\prime\right) &= \int_0^\infty dX_1 \int_0^\infty dX_2 G\left(T-t',X_{1},X_{2}\right) \\
&= \iint\limits_{\left(X_{1},X_{2}\right) \in \mathbf{D}_{12}}
G\left(\tau',X_{1},X_{2}\right) dX_{1}dX_{2}, \nonumber
\end{align}
\noindent where $\tau' =T-t'$. Similarly,
\begin{align} \label{Eq21}
q_{1}(t'&,X_1^\prime,X_2^\prime) = \iint\limits_{\left(X_{1},X_{2}\right) \in [\mathbf{D}_{12} \cup \mathbf{D}_1]} G\left(\tau',X_{1},X_{2}\right) dX_1 dX_2 \\
&+ \dfrac{1}{2}\int_{t'}^{\tau'} ds \int\limits_{\tilde{\mu}_1^<}^\infty dX_1  G_{X_{2}}\left( \tau'-s,X_{1},0\right) \chi _{1,0}\left(s,X_{1}\right) \nonumber \\
&- \int\limits_{0}^\infty dX_1  G_{X_{2}}\left( \tau'-s,X_{1},0\right) \chi _{1,\infty}\left(s,X_{1}\right)\Bigg. \nonumber
\end{align}

We start with noting that the 1D Green's function $g_{1}(\vartheta,X_{1}), \ \vartheta \equiv t-t'$ at $X_2(t) \le \mu_2^<$ has the form
\begin{equation} \label{Green1D}
g_{1}( \vartheta,X_{1}) = \dfrac{e^{-\xi _{1}^{2}\vartheta/2 + \xi_{1}\left(X_{1}-X_{1}^{\prime }\right) }}{\sqrt{2\pi \vartheta}} \left[e^{-\frac{\left(X_{1}-X_{1}^{\prime }\right)^{2}}{2\vartheta}} - e^{-\frac{\left(X_{1}+X_{1}^{\prime }-2\tilde \mu _{1}^{<}\right)^2}{2\vartheta}}\right].
\end{equation}

Accordingly,
\begin{align} \label{chi1}
&\chi_{1,0}(t',X_1^\prime) = \int \limits_{\tilde{\mu}_1^=}^\infty
e^{-\xi_1^2 \tau'/2 + \xi_1 (X_1-X_1^\prime)}
\left[\dfrac{e^{-\frac{(X_1-X_1^\prime)^2}{2\tau'}}}{\sqrt{2\pi\tau'}} - \dfrac{e^{-\frac{(X_1 + X_1^\prime - 2\tilde{\mu}_1^<)^2}{2\tau'}}}{\sqrt{2\pi\tau'}} \right]d X_1 \\
&= \int \limits_{\tilde{\mu}_1^=}^\infty
\dfrac{e^{-\frac{(X_1 - X_1^\prime - \xi_1 \tau')^2}{2\tau'}}}{\sqrt{2\pi\tau'}} dX_1
- e^{-2\xi_1 (X_1^\prime - \tilde{\mu}_1^<)}
\int \limits_{\tilde{\mu}_1^=}^\infty
\dfrac{e^{-\frac{(X_1 + X_1^\prime - 2\tilde\mu_1^< - \xi_1 \tau')^2}{2\tau'}}}{\sqrt{2\pi \tau'}} d X_1   \nonumber\\
&= N\left( -\frac{\tilde \mu_1^= -X_1^\prime - \xi_1 \tau'}{\sqrt{\tau'}}
\right) -e^{-2\xi_1(X_1^\prime - \tilde\mu_1^<) }
N\left( - \dfrac{\tilde \mu_1^= + X_1^\prime - 2\tilde\mu_1^< - \xi_1 \tau'}{\sqrt{\tau'}}
\right). \nonumber
\end{align}
\noindent and
\begin{equation} \label{chi1Inf}
\chi _{1,\infty}(t',X_1^\prime) = N\left( -\frac{\mu_1^= -X_1^\prime - \xi_1 \tau'}{\sqrt{\tau'}}\right)
-e^{-2\xi_1 X_1^\prime} N\left( - \dfrac{\mu_1^= + X_1^\prime - \xi_1 \tau'}{\sqrt{\tau'}}
\right),  \nonumber
\end{equation}

The corresponding 2D Green's function has the form (see \cite{Lipton2001,Zhou2001} and references therein)
\begin{align} \label{Green2D}
G&(\vartheta,X_1,X_2) = \\
&\dfrac{2}{\varpi \vartheta \bar{\rho}} e^{-\dfrac{\langle \xi^T,\theta \rangle \vartheta}{2} + \langle X-X^\prime, \theta \rangle - \dfrac{R^2+R^{\prime 2}}{2\vartheta}} \sum_{n=1} I_{\nu _{n}}\left( \dfrac{RR^{\prime }}{\vartheta}\right) \sin \left( \nu_n \phi \right) \sin \left( \nu _{n}\phi ^{\prime }\right), \nonumber
\end{align}
\noindent where $\langle,\rangle$ denotes the dot product, $I_k(x)$ is the modified Bessel function of the first kind,
\begin{align*}
C &=
\begin{pmatrix}
1 & \rho  \\
\rho  & 1
\end{pmatrix}
, \qquad C^{-1}= \dfrac{1}{\bar{\rho}^{2}}
\begin{pmatrix}
1 & -\rho  \\
-\rho  & 1,
\end{pmatrix}
\quad \theta = C^{-1}\xi, \quad \nu_n = \dfrac{n\pi }{\varpi},  \quad \bar\rho^2 = 1 - \rho^2, \nonumber \\
\varpi &=
\begin{cases}
\pi + \arctan\left(-\bar\rho/\rho\right), & \rho > 0 \cr
\pi/2, & \rho = 0, \cr
\arctan\left(-\bar\rho/\rho\right), & \rho < 0,
\end{cases}
\quad R^2 = \langle X, C^{-1}X^T \rangle, \quad R^{\prime 2} = \langle X^\prime, C^{-1} X^{\prime T} \rangle, \nonumber \\
&\Phi(X_1, X_2) =
\begin{cases}
\pi + \arctan\left( \dfrac{\bar{\rho}X_{1}}{-\rho X_{1} + X_{2}}\right), & X_2 < \rho X_1, \cr
\pi/2, & X_2 = \rho X_1, \cr
\arctan\left( \dfrac{\bar{\rho}X_{1}}{-\rho X_{1} + X_{2}}\right), & X_2 > \rho X_1,
\end{cases}
\nonumber \\
\phi &= \Phi(X_1,X_2), \quad \phi^\prime = \Phi(X_1^\prime, X_2^\prime), \quad
X = (X_1, X_2). \nonumber
\end{align*}

Accordingly,
\begin{align} \label{GreenDer}
G_{X_2}(\vartheta,X_1,0)  = & \dfrac{2 }{\varpi \vartheta X_1}
e^{-\dfrac{\langle \xi^T,\theta \rangle \vartheta}{2} +\theta_1 X_1 - \langle X^\prime,\theta \rangle -
\dfrac{X_1^2/\bar\rho^2+R^{\prime 2}}{2\vartheta}} \\
& \cdot \sum\limits_{n=1}\left( -1\right) ^{n+1}\nu _{n}I_{\nu _{n}}\left( \frac{%
X_{1}R^{\prime }}{\bar{\rho}\vartheta}\right) \sin \left( \nu _{n}\phi ^{\prime
}\right),  \nonumber
\end{align}

Substitution of these formulas into \eqref{Eq20}, \eqref{Eq21} yields semi-analytical expressions for $Q$ and $q_{1}$. However, from the computational point of view, it is more efficient to introduce a new function
\[ \bar{q}_1(t,X_1,X_2) = q_i(t,X_1,X_2) - \chi_{1,\infty}(t,X_1). \]
In contrast to $q_1(t,X_1,X_2)$ this new function solves a problem similar the problem given in \eqref{Eq13}, but with a homogeneous upper boundary condition:
\begin{eqnarray} \label{Eq13new}
\bar{q}_{1,t}(t,X_1,X_2) + \mathcal{L}\bar{q}_1(t,X_1,X_2) &=& 0, \\
\bar{q}_1(t,0,X_2) &=& \bar{q}_1(t,X_1,X_2\uparrow\infty) = 0, \nonumber \\
\bar{q}_1(T,X_1,X_2) &=& -\mathbf{1}_{X \in \bar{\mathbf{D}}},   \nonumber
\end{eqnarray}
\noindent where $\bar{\mathbf{D}}$ is the area inside the curvilinear triangle with the  vertexes in points  6-7-9 in Fig.~\ref{Fig2}.

As the equations in \eqref{Eq13} and \eqref{Eq13new} differ just by the source term, the Green's function of \eqref{Eq13new} is also given by \eqref{Green2D}. Accordingly, the solution of \eqref{Eq13new} reads
\begin{align} \label{Eq21new}
q_1(t'&,X_1^\prime,X_2^\prime) = \chi_{1,\infty}(t',X^\prime_1) -
\iint\limits_{(X_1,X_2) \in \bar{\mathbf{D}}} G\left(\tau',X_{1},X_{2}\right) dX_1 dX_2 \\
&+ \dfrac{1}{2} \int_{0}^{\tau'} d s \int_0^\infty G_{X_2}(\tau'-s,X_1,0)
\left[\Xi(\tau'-s,X_1) - \chi_{1,\infty}(\tau'-s,X_1)\right] d X_1.  \nonumber
\end{align}

Another simplification could be made if one wishes to compute the first integral in \eqref{Eq21}. For a better accuracy it could be reasonable to represent it as a difference of two integrals. The first one is taken over the positive quadrant $(X_1,X_2) \in [0,\infty)\times [0,\infty)$ while the second integral is defined in a union of two semi-infinite strips: $\mathcal{D}_1 \cup \mathcal{D}_2 \cup \mathcal{D}$. The idea is that the second integral is defined in the area which is finite either in one or the other direction, and the first integral could be represented in the closed form. Therefore, the total computational error is less. We underline, that to the best of our knowledge representation of the first integral in closed form yet was not given in the literature, so we present this derivation in Appendix~\ref{App1}.

\subsection{Numerical experiments}
In our test examples we solved \eqref{Eq13} by using first a finite difference scheme (FD) and then compared it with the analytical solution given by \eqref{Eq21new}. Since \eqref{Eq13} is a pure convection-diffusion two-dimensional problem, we solved it numerically by using a Hundsdorfer-Verwer scheme, see \cite{HoutFoulon2010}. A non-uniform finite-difference grid was constructed similar to \cite{ItkinCarrBarrierR3} with the grid nodes concentrated close to $\tilde\mu^=_i, \ i=1,2$. We solved the problem using parameters given in Table~\ref{Tab2}\footnote{In our setting the value of the interest rate $r$ doesn't matter.}:
\begin{table}[!ht]
\begin{center}
\caption{Parameters of the structural default model.}
\begin{tabular}{|c|c|c|c|c|c|c|c|c|c|}
\hline
$L_{1,0}$ & $L_{2,0}$ & $L_{12,0}$ & $L_{21,0}$ & $R_{1}$ & $R_{2}$ & T & $\sigma_1$ & $\sigma_2$ & $\rho$ \cr
\hline
60 & 70 & 10 & 15 & 0.4 & 0.45 & 1 & 1 & 1 & 0.5 \cr
\hline
\end{tabular}
\label{Tab2}
\end{center}
\end{table}

\begin{figure}[!ht]
\begin{minipage}{0.46\linewidth}
\begin{center}
\fbox{\includegraphics[width=\linewidth]{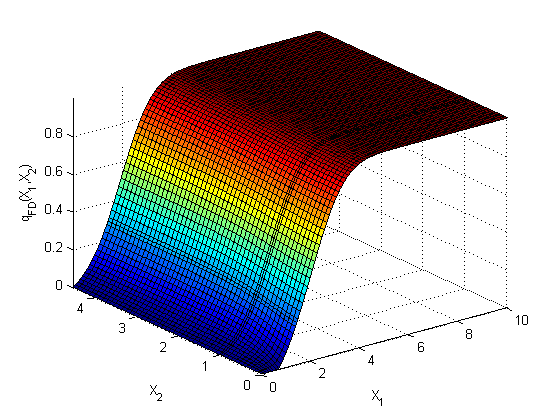}}
\caption{The marginal survival probability $q_1(X_1,X_2)$ computed by using a Hundsdorfer-Verwer scheme. \hfill \break}
\label{q1FD}
\end{center}
\end{minipage}
\hspace{0.04\linewidth}
\begin{minipage}{0.46\linewidth}
\begin{center}
\fbox{\includegraphics[width=\linewidth]{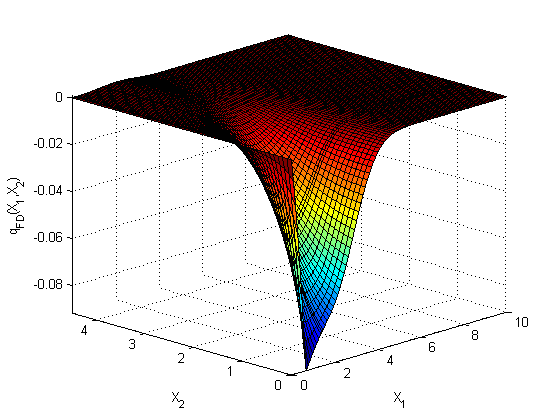}}
\caption{The difference between marginal survival probabilities $q_1(X_1,X_2)$ computed with and without mutual obligations using the FD method.}
\label{FigDifFD}
\end{center}
\end{minipage}
\end{figure}

We computed all tests using a $100 \times 100$ spatial grid. Also we used a constant step in time $\Delta \tau = 0.01$, so that the total number of time steps for a given maturity is 100.
The marginal survival probability $q_1(X_1,X_2)$ at $t=0$ computed by using this method is presented in Fig.~\ref{q1FD}.

It is easy to see that for our chosen parameters $\tilde{\mu}^<_1 = 0.6659, \ \tilde{\mu}^<_2 = 0.2548, \
\mu^=_1 = 1.4424, \ \mu^=_2 = 0.9764, \ \tilde{\mu}^=_1 = 1.5821, \ \tilde{\mu}^=_2 = 1.0534$.

To observe the effect of the mutual liabilities we repeated this test, but with zero mutual liabilities. 
Therefore, as compared with the previous case, now the total assets of the $i$-th bank are $A_i + \sum_{j} L_{ij}$ while its liabilities are $L_i + \sum_{j} L_{ji}$. But to provide a correct comparison we need to keep the asset values $A_i$ constant. Therefore, in this case we re-adjust liabilities to $L_i + \sum_{j} L_{ji} - \sum_{j} L_{ij}$. In words, that means that, if $\sum_{j} L_{ij}$ is positive, the bank $i$ gets extra cash and then spends it retiring some of its external liabilities. If this amount is negative, then it is borrowed from the external sources. After this adjustment is done we set $L_{21} = L_{12} = 0$ in our calculations. In what follows we call this procedure an Adjustment Procedure (AP).

The difference of two solutions is presented in Fig.~\ref{FigDifFD}. The above plot clearly demonstrates a significant difference in the solution in the area close to $X_1 = \mu^=_1, \ X_2 = \mu^=_2$, i.e. the effect of the mutual liabilities is pronounced in this area.

\begin{figure}[!ht]
\begin{minipage}{0.46\linewidth}
\begin{center}
\fbox{\includegraphics[width=\linewidth]{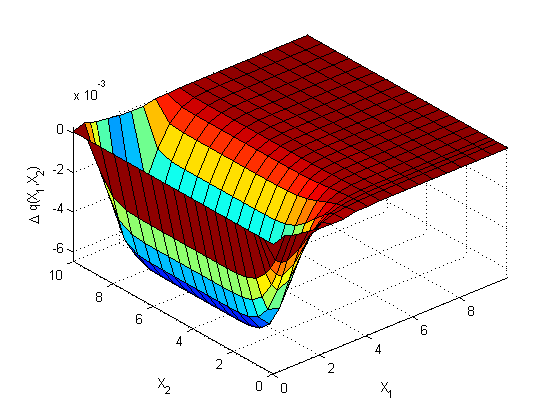}}
\caption{The difference between marginal survival probabilities $q_1(X_1,X_2)$ computed by the analytical and FD methods with no mutual obligations. }
\label{FD_AnalNMO}
\end{center}
\end{minipage}
\hspace{0.04\linewidth}
\begin{minipage}{0.46\linewidth}
\begin{center}
\fbox{\includegraphics[width=\linewidth]{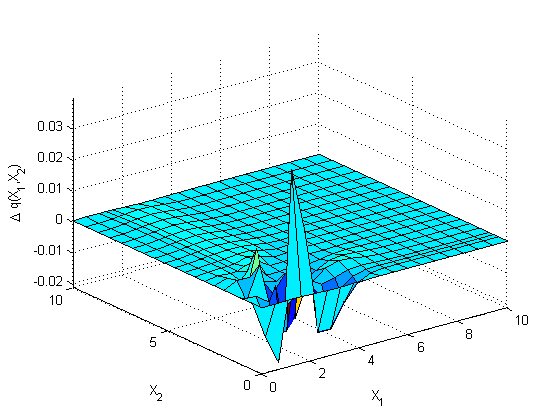}}
\caption{The difference between marginal survival probabilities $q_1(X_1,X_2)$ computed by the analytical and FD methods with mutual obligations, $T=5$ years.}
\label{q1FD_Anal}
\end{center}
\end{minipage}
\end{figure}

Next we want to compare the analytical and FD solutions. Since the integrands in \eqref{Eq21new} are highly oscillating functions, to get a reasonable accuracy we used a Gauss-Kronrod algorithm in both directions. Fig.~\ref{FD_AnalNMO} demonstrates the difference in these solutions when there are no mutual obligations.
Both solutions coincide pretty well. However, when mutual obligations are taken into account the difference increases as it can be seen in Fig.~\ref{q1FD_Anal}. The difference is bigger in the area closer to $X_1 = \mu^=_1$.

Performance-wise, computation of the marginal probabilities on this spatial grid using the FD scheme takes 21 secs for $T=1$ year. At the same time, computation of a single point on a grid using our analytical methods takes 0.4-0.6 secs\footnote{We ran this test in Matlab on a standard PC with Intel Xeon E5620 2.4 Ghz CPU.}. Therefore, if the marginal probabilities should be computed at every node of the FD grid, using the analytical method it would take about 4000 secs, which is pretty slow. However, when calibrating the model with unknown volatilities and the correlation coefficient, we need functions at only three quotes that could be taken from the market values of the CDS spreads and First-To-Default swaps spreads. Thus, in our simplistic model we need just three points, which takes about 1.2 secs to compute using our analytics. In contrast, the FD scheme cannot be reduced just to three points on the grid, and, therefore, for such kind of calibration is much less efficient than the analytical method. This is the reason we propose the approach of this paper for doing fast calibration of the model.

In a more general setting, e.g., the one proposed in \cite{ItkinLipton2014} this simplified approach could be used to produce a "smart" initial guess for the parameters of the marginal distributions. Then, using this guess, the whole rather complicated problem could be calibrated much faster than starting with some arbitrary values of the parameters, since in this case only relatively small increments of the initial guess should be found.

\section{Pricing CDS contracts} \label{sCDS}

We now describe how to price CDSs and FTD in our setting.
\subsection{CDS}
The price of a CDS $C_1(t,X_1,X_2)$\footnote{For $C_2(t,X_1,X_2)$ similar expressions could be provided by analogy.} written on the first bank solves the following problem, (\cite{Bielecki2004}):
\begin{align} \label{cds}
C_{1,t}(t,X_1,X_2) &+ \mathcal{L} C_1(t,X_1,X_2) = \varsigma_1,   \\
C_1(t,X_1,0) &= \Psi(t,X_1) =
\begin{cases}
c_{1,0}(t,X_1), & X_1 > \tilde{\mu}_1^< \\
1 - R_1, & X_1 \le \tilde{\mu}_1^<
\end{cases}
, \nonumber \\
C_1(t,0,X_2) &= 1-R_1,  \quad C_1(t,X_1,X_2\uparrow\infty) = c_{1,\infty}(t,X_1),
\nonumber
\end{align}
\noindent where $\varsigma_i$ is the coupon rate, $c_{1,0}(t,X_1)$ is the solution of the one-dimensional terminal boundary value problem
\begin{align} \label{c10}
\partial_t c_{1,0}(t,X_1) &+ \mathcal{L}_1 c_{1,0}(t,X_1) = \varsigma_1, \\
c_{1,0}(t,\tilde{\mu}_1^<) &= 1 - R_1, \quad c_{1,0}(t,\infty) = -\varsigma_1 (T-t), \nonumber \\
c_{1,0}(T,X_1) &= (1-R_1) \mathbf{1}_{\tilde{\mu}_1^< \le X_1 \le \tilde \mu_1^=}, \nonumber
\end{align}
\noindent and $c_{1,\infty}(t,X_1)$ is the solution of another one-dimensional terminal boundary value problem
\begin{align} \label{c10}
\partial_t c_{1,\infty}(t,X_1) &+ \mathcal{L}_1 c_{1,\infty}(t,X_1) = \varsigma_1, \\
c_{1,\infty}(t,0) &= 1 - R_1, \quad c_{1,\infty}(t,\infty) = -\varsigma_1 (T-t), \nonumber \\
c_{1,\infty}(T,X_1) &= (1-R_1) \mathbf{1}_{X_1 \le \mu_1^=}. \nonumber
\end{align}

Also the statement of problem given in \eqref{cds} must be supplied with the terminal condition $C_1(T,X_1,X_2)$, which could be provided based on the picture presented in Fig.~\ref{Fig2}. Omitting some intermediate algebra, we obtain the following condition
\begin{align*}
C_1(T,X_1,X_2) &= \alpha_1 \mathbf{1}_{(X_1,X_2) \in [\hat{\mathbf{D}} \cup \mathbf{D}_2]} \\
\alpha_1(X_1,X_2) &=
\begin{cases}
1- \min[\tilde R_{1,T}(1), R_1], & (X_1, X_2) \in \mathbf{D}_{2} \cr
1- \min[\tilde R_{1,T}(\gamma_2), R_1], &  (X_1, X_2) \in \hat{\mathbf{D}}.
\end{cases}
\nonumber
\end{align*}

The value of the components $\gamma_{i}$ at $(X_1, X_2) \in \hat{\mathbf{D}}$ are determined by solving the detailed balance equations which follow from the general $N$-dimensional problem given in \eqref{gamProblem}
\begin{align*}
A_{1}+\gamma _{2}L_{21} &= \gamma _{1}\left( L_{1}+L_{12}\right),  \\
A_{2}+\gamma _{1}L_{12} &= \gamma _{2}\left( L_{2}+L_{21}\right),  \nonumber
\end{align*}
The solution in the original variables reads
\[ \gamma_i = \dfrac{L_{\bar{i}}A_{i}+L_{\bari,i}(A_{i}+A_{\bari})}{\bigtriangleup}, \quad i=1,2. \]

Observe that the Green's function for \eqref{c10} is that given by \eqref{Green1D}. Therefore,
\begin{align} \label{c10sol}
c_{1,0}(t^\prime,X_1^\prime) &= (1-R_1)\int_{\tilde{\mu}_1^<}^{\tilde\mu_{1}^=} g_{1}( \tau^\prime,X_{1}) d X_1 + \dfrac{1-R_1}{2} \int_{0}^{\tau^\prime} \fp{g_{1}( \tau^\prime-s,X_{1})}{X_1}\Bigg|_{X_1=\tilde{\mu}_1^<} d s \nonumber \\
&- \varsigma_1 \int_{0}^{\tau^\prime} \int_{\tilde\mu_{1}^<}^\infty  g_{1}( \tau^\prime-s,X_{1}) d X_1 d s
\equiv I_1 + I_2 + I_3.
\end{align}

All these integrals can be computed in the closed form. Omitting some intermediate algebra we provide just the final results:
\begin{align} \label{I123}
I_1 &= (1-R_1)\Bigg\{e^{- 2 \xi_1 (X_1' - \tilde{\mu}_1^<)}
\left[N\left(-y_-\right) - N\left(-2 y_- - z\right) \right]
+ N\left(y_+\right) - N\left(z\right)\Bigg\}, \nonumber \\
I_2 &= (1-R_1)\left[e^{-2 \xi_1  (X_1^\prime - \tilde{\mu}_1^< )}
N\left(y_-\right) + N\left(-y_+\right)\right], \\
I_3 &= -\varsigma_1 \tau' \left[1 - \frac{y_+ }{\xi_1 \sqrt{\tau'}} N\left(-y_+\right)
- e^{-2 \xi_1 (X_1^\prime - \tilde{\mu}_1^<)}
\frac{y_- }{\xi_1 \sqrt{\tau'}} N\left(y_-\right) \right], \nonumber \\
y_\pm &= \dfrac{\pm(X_1^\prime-\tilde{\mu}_1^<) + \xi_1 \tau'}{\sqrt{\tau'}}, \quad
z = \dfrac{X_1^\prime -\tilde{\mu}_1^= + \tau' \xi_1}{\sqrt{\tau'}}. \nonumber
\end{align}

By analogy
\begin{align} \label{c1infsol}
c_{1,\infty}(t^\prime,X_1^\prime) &= (1-R_1)\int_0^{\mu_{1}^=} \bar{g}_{1}( \tau^\prime,X_{1}) d X_1 + \dfrac{1-R_1}{2} \int_{0}^{\tau^\prime} \fp{\bar{g}_{1}( \tau^\prime-s,X_{1})}{X_1}\Bigg|_{X_1=0} d s \nonumber \\
&- \varsigma_1 \int_{0}^{\tau^\prime} \int_0^\infty  \bar{g}_{1}( \tau^\prime-s,X_{1}) d X_1 d s,
\end{align}
\noindent where $\bar{g}_{1}( \tau^\prime,X_{1})$ can be obtained from $g_{1}( \tau^\prime,X_{1})$ by setting in \eqref{Green1D} $\tilde{\mu}_1^< = 0$. Accordingly, these integrals in closed form are given by \eqref{I123} by replacing $\tilde{\mu}_1^< = 0$ and $\tilde{\mu}_1^= = \mu_1^=$.

Using the same trick as in the previous section when we computed the marginal survival probability $q_1(t,X_1,X_2)$, the final solution of this problem could be represented as follows:
\begin{align} \label{CDSnew}
C_1&(t',X_1^\prime,X_2^\prime) = c_{1,\infty}(t',X^\prime_1) + \int_0^\infty \int_0^\infty \phi(X_1,X_2) G\left(\tau',X_{1},X_{2}\right) dX_1 dX_2 \\
&+ \dfrac{1}{2} \int_{0}^{\tau'} d s \int_0^\infty G_{X_2}(\tau'-s,X_1,0)
\left[ \Psi(\tau'-s,X_1) - c_{1,\infty}(\tau'-s,X_1)\right] d X_1, \nonumber \\
&\phi(X_1,X_2) \equiv \alpha_1 \mathbf{1}_{(X_1,X_2) \in [\hat{\mathbf{D}} \cup \mathbf{D}_2]} -(1-R_1) \mathbf{1}_{X_1 < \mu_1^=}, \nonumber
\end{align}
\noindent where the Green's function $G\left(\tau',X_{1},X_{2}\right)$ is given in \eqref{Green2D}.

\subsection{Numerical experiments}
It is well-known in a theory of heat conduction that a direct implementation of \eqref{CDSnew} is still impractical. The reason is that at $X_2^\prime = 0$ the first integral in \eqref{CDSnew} vanishes, so the second one must converge to $c_{1,0}(t',X^\prime_1) - c_{1,\infty}(t',X^\prime_1)$ to provide the correct boundary condition at $X_2^\prime = 0$. However, as it could be checked, at $X_2^\prime = 0$ we have $G_{X_2}(\tau'-s,X_1,0) = 0$, and, hence, the formal validation of \eqref{Eq21} fails at the boundary. It is explained, e.g. in \cite{Kartashov2001}, the reason is that the series in the second integral in \eqref{CDSnew} is not uniformly convergent at $X_2^\prime = 0$, so the transition to the limit $X_2^\prime \to 0$ using this representation is complicated and impractical from the computational point of view.

This problem, however, can be overcome by applying another elegant trick that we describe in more detail in Appendix~\ref{App2}.

\begin{figure}[!ht]
\begin{minipage}{0.46\linewidth}
\begin{center}
\fbox{\includegraphics[width=\linewidth]{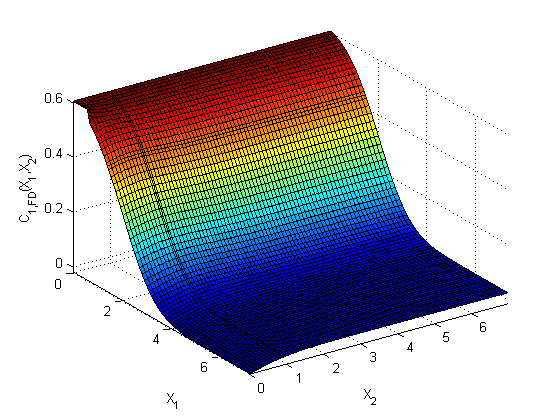}}
\caption{CDS prices $C_1(X_1,X_2)$ computed by using a Hundsdorfer-Verwer scheme.
\hfill \break}
\label{cdsFD}
\end{center}
\end{minipage}
\hspace{0.04\linewidth}
\begin{minipage}{0.46\linewidth}
\begin{center}
\fbox{\includegraphics[width=\linewidth]{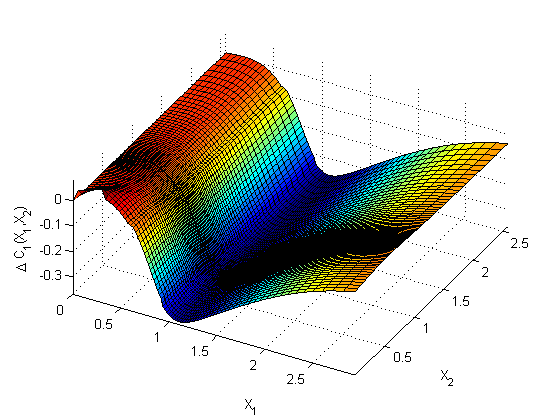}}
\caption{The difference in CDS prices $C_1(X_1,X_2)$ computed with and without mutual obligations using the FD method.}
\label{cdsDifWith-WithoutFD}
\end{center}
\end{minipage}
\end{figure}

Further on, we ran the same test as in the previous section with parameters of the model given in Table~\ref{Tab2}, and $\varsigma_1 = 0.02$. We used the same FD method to verify our solution, see the previous section for the description of the method. The CDS prices for $t=0$ are presented in Fig.~\ref{cdsFD}.

Again, to observe the effect of mutual liabilities we perform an equivalent computations, but with zero mutual liabilities and the AP applied. The difference of two solutions is presented in Fig.~\ref{cdsDifWith-WithoutFD}. As one would expect, our results demonstrate a significant difference in the area close to $X_1 = \mu^=_1, \ X_2 = \mu^=_2$, i.e. the effect of the mutual liabilities is pronounced in this area not only for the marginal survival probabilities, but for CDS prices as well.

\section{Pricing first-to-default (FTD) contracts} \label{sFTD}

The price of the FTD $F_{1,t}$ solves the following terminal boundary value problem\footnote{For $F_{2,t}$ this can be done by analogy.}:
\begin{align} \label{FTD}
F_{1,t} + \mathcal{L}_1 F_1 &= \varsigma_1,  \\
F_1(t,0,X_2) &= 1-R_1, \qquad F_1(t,X_1,0) = 1-R_2.  \nonumber \\
F_1(t,X_1,X_2\uparrow\infty) &= f_{1,\infty}(t,X_1), \quad
F_1(t,X_1\uparrow\infty, X_2) = f_{2,\infty}(t,X_2), \nonumber \\
F_1(T,X_1,X_2) &=
\beta_0 \mathbf{1}_{(X_1,X_2) \in \hat{\mathbf D}} +
\beta_1 \mathbf{1}_{(X_1,X_2) \in \mathbf{D}_1} +
\beta_2 \mathbf{1}_{(X_1,X_2) \in \mathbf{D}_2}. \nonumber
\end{align}

Here $f_{i,\infty}(t,X_i), \ i=1,2$ is the solution of the one-dimensional terminal boundary value problem
\begin{align} \label{f10}
\partial_t f_{i,\infty}(t,X_i) &+ \mathcal{L}_i f_{i,\infty}(t,X_i) = \varsigma_i, \\
f_{i,\infty}(t,0) &= 1 - R_i, \quad f_{i,\infty}(t,\infty) = -\varsigma_i (T-t), \nonumber \\
f_{i,\infty}(T,X_i) &= (1-R_i) \mathbf{1}_{X_i \le \mu_i^=}. \nonumber
\end{align}
As it could be seen, $f_{1,\infty}(t,X_1) = c_{1,\infty}(t,X_1)$ given in \eqref{c1infsol}. Also $\beta_i = \beta_i(X_1,X_2)$ in \eqref{FTD} is defined as
\begin{align*}
\beta_i &= 1- \min[\tilde R_{\bari,T}(1), R_{\bari}], \quad  (X_1, X_2) \in \mathbf{D}_i, \ i=1,2, \\
\beta_0 &=  1- \min[ \min[\tilde R_{2,T}(\gamma_1), R_{2}], \min[\tilde R_{1,T}(\gamma_2), R_{1}]], \quad (X_1, X_2) \in \hat{\mathbf D}.
\end{align*}

Similar to the previous section, it can be shown that the solution of this problem reads
\begin{align} \label{FTDnew}
F_1&(t',X_1^\prime,X_2^\prime) = \int_0^\infty \int_0^\infty \beta_1 \mathbf{1}_{(X_1,X_2) \in [\hat{\mathbf{D}} \cup \mathbf{D}_2]} G\left(\tau',X_{1},X_{2}\right) dX_1 dX_2 \\
&- \varsigma_1 \int_{0}^{\tau^\prime} \int_0^\infty  \bar{g}_{1}( \tau^\prime-s,X_{1}) d X_1 d s
+ \dfrac{1}{2}(1-R_2) \int_{0}^{\tau'} d s \int_0^\infty G_{X_2}(\tau'-s,X_1,0)d X_1 \nonumber \\
&+ \dfrac{1}{2}(1-R_1) \int_{0}^{\tau'} d s \int_0^\infty G_{X_1}(\tau'-s,0,X_2)d X_2, \nonumber
\end{align}
\noindent where the Green's function $G\left(\tau',X_{1},X_{2}\right)$ again is given in \eqref{Green2D}.

\subsection{Numerical experiments}
For the same reason as before a direct implementation of \eqref{FTDnew} is impractical from the computational point of view. However, again a similar trick can be applied to significantly improve the accuracy in computation of the boundary integrals. We describe it in more detail in Appendix~\ref{App3}.

\begin{figure}[!ht]
\begin{minipage}{0.46\linewidth}
\begin{center}
\fbox{\includegraphics[width=\linewidth]{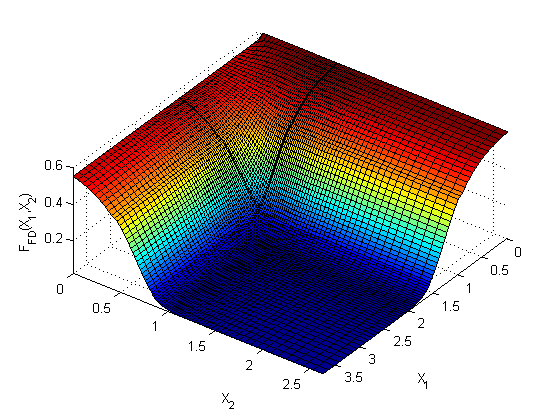}}
\caption{FTD prices $F_1(X_1,X_2)$ computed by using a Hundsdorfer-Verwer scheme.
\hfill \break}
\label{ftdFD}
\end{center}
\end{minipage}
\hspace{0.04\linewidth}
\begin{minipage}{0.46\linewidth}
\begin{center}
\fbox{\includegraphics[width=\linewidth]{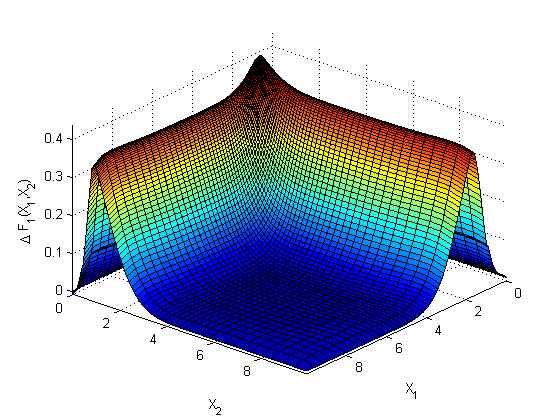}}
\caption{The difference in FTD prices $F_1(X_1,X_2)$ computed without and with mutual obligations using the FD method.}
\label{ftdFDnmo-FD}
\end{center}
\end{minipage}
\end{figure}

Again we ran the same test as in the previous section with parameters of the model given in Table~\ref{Tab2}, and the same $\varsigma_1 = 0.02$. We used the same FD method to verify our solution, see the previous section for the description of the method. The FTD prices are shown in Fig.~\ref{ftdFD}.

In order to understand the effect of the mutual liabilities on FTD prices, we repeated this test, but with zero mutual liabilities and the AP applied. The difference of two solutions is presented in Fig.~\ref{ftdFDnmo-FD}. Same picture can be obtained by using our analytical approach.

As in the previous cases mutual obligations significantly influence FTD prices especially in the area close to $X_1 = \mu^=_1, \ X_2 = \mu^=_2$.

\begin{figure}[!ht]
\begin{minipage}{0.46\linewidth}
\begin{center}
\fbox{\includegraphics[width=\linewidth]{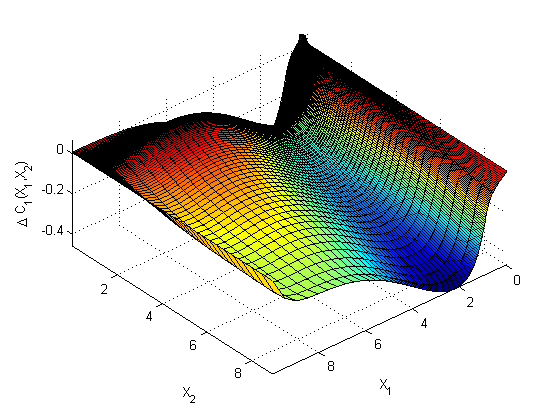}}
\caption{The difference in CDS prices $C_1(X_1,X_2)$ computed with and without mutual obligations, T=5 years.}
\label{cdsMO-NMO_5}
\end{center}
\end{minipage}
\hspace{0.04\linewidth}
\begin{minipage}{0.46\linewidth}
\begin{center}
\fbox{\includegraphics[width=\linewidth]{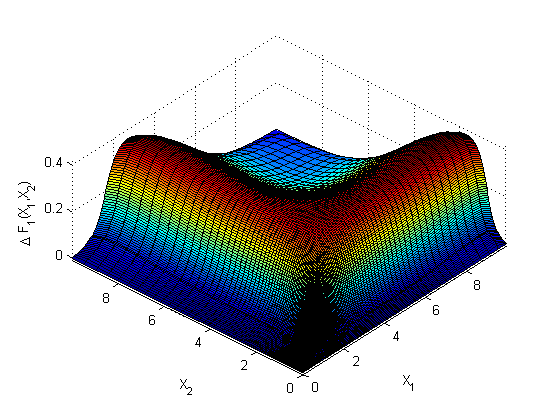}}
\caption{The difference in FTD prices $C_1(X_1,X_2)$ computed without and with mutual obligations, T=5 years.}
\label{ftdNMO-MO_5}
\end{center}
\end{minipage}
\end{figure}

We also present the difference in the CDS and FTD prices for the first bank computed  with and without mutual obligations and maturity $T=5$ years. These results are given in Fig.~\ref{cdsMO-NMO_5}, \ref{ftdNMO-MO_5}. It is seen that with the increase of maturity the effects of the mutual obligations decreases and in the limit of very long maturities almost disappears.

\begin{figure}[!ht]
\begin{center}
\fbox{\includegraphics[width=0.5\linewidth]{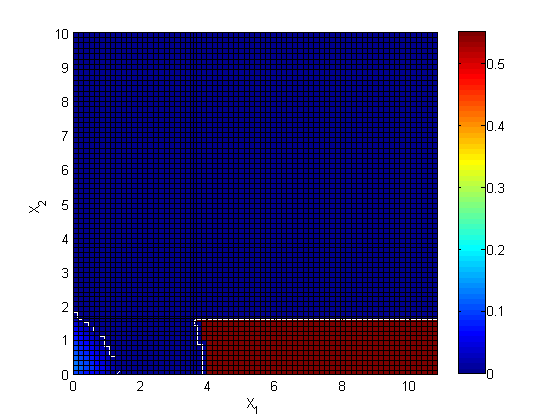}}
\caption{The difference $F_1(T,X_1,X_2) - C_1(T,X_1,X_2)$, T=1 year.}
\label{ftdFDnmo-cdsT}
\end{center}
\end{figure}

To illustrate how the terminal distribution of prices looks like in some particular example (which was schematically given in Fig.~\ref{Fig2}) in Fig.~\ref{ftdFDnmo-cdsT} the difference $F_1(T,X_1,X_2) - C_1(T,X_1,X_2)$ is presented as a function of $(X_1,X_2)$. Obviously,
$F_1(T,X_1,X_2)$ is positive in $\mathbf{D}_2$ while $C_1(T,X_1,X_2)$ vanishes there (the red box in the right bottom corner of the Figure). And they also differ in a part of the domain $\tilde{\mathbf{D}}$ (the left bottom corner). In other points of the computational domain the values of $F_1(T,X_1,X_2)$ and $C_1(T,X_1,X_2)$ coincide with each other.

\section{Calibration} \label{Calib}
The model described in section~\ref{ge} has three unknown parameters: $\sigma_1, \sigma_2, \rho$. We use CDS prices for two assets and the FTD price for both assets to calibrate these parameters. The calibration is done in Matlab using a simple non-linear least square approach where every given point (quote) is taken with the same weight.

In a test experiment we use all the parameters as in Table~\ref{Tab2} and also use $A_{1,0} = 300, \ A_{2,0} = 300, \ \varsigma_1 = \varsigma_2 = 0.05$. Then we calibrate $\sigma_1, \sigma_2, \rho$ in the following way. First, we set $\sigma_1 = 0.3, \sigma_2 = 0.4, \rho = 0.5$ and compute the prices of CDS and FTD using our algorithm. This gives us the quotes $C_1 = 0.05, \ C_2 = 0.0583, \ F_1 = 0.0583$. Then we run the calibrator to make sure it converges to the same values of $\sigma_1, \sigma_2, \rho$ to validate self-consistence of our approach.

Finally, since we investigate how strong the effect of mutual obligations on the parameters of the model is, in the second test we ignore mutual obligations and apply our AP as was discussed in the previous sections. The results of such a calibration are presented in Table~\ref{calib}.
\begin{table}[!ht]
\begin{center}
\caption{Results of calibration}
\begin{tabular}{|c|c|c|c|c|c|c|}
\hline
  & \multicolumn{3}{|c|}{T = 1 yr} & \multicolumn{3}{|c|}{T = 5 yrs} \\
\hline
Test & $\sigma_1$ & $\sigma_2$ & $\rho$ & $\sigma_1$ & $\sigma_2$ & $\rho$ \\
\hline
MO & 0.300 & 0.400 & 0.500 & 0.300 & 0.400 & 0.500 \\
\hline
NMO &  0.2819 & 0.4421 & 0.4936 & 0.3189 & 0.4234 & 0.2942  \\
\hline
Dif, \% & 6.0372 & -10.5373 &  1.2801 & -6.3108  & -5.8616  & 41.1670  \\
\hline
\end{tabular}
\label{calib}
\end{center}
\end{table}

A typical time necessary to get the values of the parameters in Matlab is about 10 secs with the objective function tolerance set to $10^{-4}$. The corresponding time if the FD algorithm is used with the grid $70 \times 70$ points in space and time step 0.03 is about 12 times slower. Certainly, for longer maturities this difference increases. The results for $T=5$ years are also presented in Table~\ref{calib}. Here the computed quotes are $C_1 = 0.2579, \ C_2 = 0.3182, \ F_1 = 0.336$. As it can be seen, accounting for the mutual obligations significantly affects the values of the calibrated parameters.

\section{Conclusion} \label{Concl}
In this paper we consider interlinkage (mutual obligations) of banks and their influence on marginal survival probabilities as well as CDS and FTD prices of the corresponding names. We use a simple model where banks' assets are driven by the correlated Brownian motions with drift. The choice of the model is dictated by the advantage to get all the results in a closed form, at least in the 2D case. A more sophisticated model withs assets driven by a general correlated L\'{e}vy processes has been already considered in \cite{ItkinLipton2014}. However, the present description is more transparent and allows one to better understand the nature of the effect, and also adds CDS and FTD prices to the picture. In the 2D case we also calibrated this model to some artificial market quotes and showed that the mutual obligations must be taken into account to get the correct values of the model parameters as they significantly influence the results of calibration. To the best of our knowledge these results are new.

Another less important, but perhaps interesting result is a closed form solution for the marginal survival probabilities for two assets driven by the correlated Brownian motions with drift. This solution yet was not given in the literature, so we present it in this paper. To understand a financial meaning of this solution, we underline that for big banks due to various regulators requirements their assets cannot drop down much below their liabilities, which means that their recovery rates $R$ should be almost 1 (or, possibly, even exceed 1). In this case when computing, e.g., marginal survival probabilities, the domain $\mathbf{D}_{12}$ in Fig.~\ref{Fig2} becomes a positive octant.

 \section*{Acknowledgments}
We thank Darrel Duffie, Dilip Madan and Tore Opsahl for useful comments. We assume full responsibility for any remaining errors.
 

\begin{thebibliography}{10}
\providecommand{\url}[1]{{#1}}
\providecommand{\urlprefix}{URL }
\expandafter\ifx\csname urlstyle\endcsname\relax
  \providecommand{\doi}[1]{DOI~\discretionary{}{}{}#1}\else
  \providecommand{\doi}{DOI~\discretionary{}{}{}\begingroup
  \urlstyle{rm}\Url}\fi

\bibitem{as64}
Abramowitz, M., Stegun, I.: Handbook of Mathematical Functions.
\newblock Dover Publications, Inc. (1964)

\bibitem{Abreu2008}
Abreu, L.: The reproducing kernel structure arising from a combination of
  continuous and discrete orthogonal polynomials into fourier systems (2008).
\newblock Available at \url{http://arxiv.org/abs/math/0601190}

\bibitem{Askey2010}
Askey, R., Daalhuis, A.B.O.: Generalized hypergeometric function.
\newblock In: F.~Olver, D.~Lozier, R.~Boisvert, C.~Clark (eds.) NIST Handbook
  of Mathematical Functions. Cambridge University Press (2010)

\bibitem{Bellman}
Bellman, R.: Introduction to matrix analysis.
\newblock McGraw--Hill (1960)

\bibitem{Bielecki2004}
Bielecki, T.R., Rutkowski, M.R.: Credit Risk: Modeling, Valuation and Hedging.
\newblock Springer (2004)

\bibitem{BC1976}
Black, F., Cox, J.: Valuing corporate securities: Some effects of bond
  indenture provisions.
\newblock Journal of Finance \textbf{31}(2), 351--367 (1976)

\bibitem{Patras2008}
Blanchet-Scalliet, C., Patras, F.: Counterparty risk valuation for {CDS}
  (2008).
\newblock Available at \url{http://arxiv.org/abs/0807.0309}

\bibitem{DavidLehar2014}
David, A., Lehar, A.: Why are banks highly interconnected? (2014).
\newblock Available at \url{ http://dx.doi.org/10.2139/ssrn.1108870 }

\bibitem{EN2001}
Eisenberg, L., Noe, T.: Systemic risk in financial systems.
\newblock Management Science \textbf{47}, 236--249 (2001)

\bibitem{GR2007}
Gradshtein, I., Ryzhik, I.: Table of Integrals, Series, and Products.
\newblock Elsevier (2007)

\bibitem{FPT2003}
Granas, A., Dugundji, J.: Fixed point theory.
\newblock Springer Verlag, New York (2003)

\bibitem{HoutFoulon2010}
{In't Hout}, K.J., Foulon, S.: {ADI} finite difference schemes for option
  pricing in the {H}eston model with correlation.
\newblock International journal of numerical analysis and modeling
  \textbf{7}(2), 303--320 (2010)

\bibitem{Itkin2014}
Itkin, A.: Efficient solution of backward jump-diffusion {PIDEs} with splitting
  and matrix exponentials.
\newblock Journal of Computational Finance \textbf{Forthcoming} (2014).
\newblock Electronic version is available at
  \url{http://arxiv.org/abs/1304.3159}

\bibitem{ItkinCarrBarrierR3}
Itkin, A., Carr, P.: Jumps without tears: A new splitting technology for
  barrier options.
\newblock International Journal of Numerical Analysis and Modeling
  \textbf{8}(4), 667--704 (2011)

\bibitem{ItkinLipton2014}
Itkin, A., Lipton, A.: Efficient solution of structural default models with
  correlated jumps and mutual obligations (2014).
\newblock Availbale at \url{http://arxiv.org/abs/1408.6513}

\bibitem{Kartashov2001}
Kartashov, E.: Analytical Methods in the Theory of Heat Conduction in Solids.
\newblock Vysshaya Shkola, Moscow (2001)

\bibitem{Kythe2011}
Kythe, P.: Green's Functions and Linear Differential Equations: Theory,
  Applications, and Computation.
\newblock Applied Mathematics \& Nonlinear Science. Chapman \& Hall/CRC (2011)

\bibitem{Lipton2001}
Lipton, A.: Mathematical Methods For Foreign Exchange: A Financial Engineer's
  Approach.
\newblock World Scientific (2001)

\bibitem{LiptonSavescu2014}
Lipton, A., Savescu, I.: Pricing credit default swaps with bilateral value
  adjustments.
\newblock Quantitative Finance \textbf{14}(1), 171--188 (2014)

\bibitem{LiptonSepp2011}
Lipton, A., Sepp, A.: Credit value adjustment in the extended structural
  default model.
\newblock In: The Oxford Handbook of Credit Derivatives, pp. 406--463. Oxford
  University (2011)

\bibitem{m74}
Merton, R.: On the pricing of corporate debt: The risk structure of interest
  rates.
\newblock Journal of Finance \textbf{29}, 449---470 (1974)

\bibitem{Metzler2010}
Metzler, A.: On the first passage problem for correlated {Brownian} motion.
\newblock Statistics and Probability Letters \textbf{80}, 277--284 (2010)

\bibitem{Polyanin2002}
Polyanin, A.: Handbook of linear partial differential equations for engineers
  and scientists.
\newblock Chapman \& Hall/CRC (2002)

\bibitem{Watson1966}
Watson, G.: A Treatise on the Theory of Bessel Functions, 2nd edn.
\newblock Cambridge University Press, Cambridge, UK (1966)

\bibitem{BOE2011}
Webber, L., Willison, M.: Systemic capital requirements.
\newblock Tech. Rep. 436, Bank of England (2011).
\newblock Available at
  \url{http://papers.ssrn.com/sol3/papers.cfm?abstract_id=1945654}

\bibitem{Zhou2001}
Zhou, C.: The term structure of credit spreads with jump risk.
\newblock Journal of Banking and Finance \textbf{25}, 2015--2040 (2001)

\end{thebibliography}

\newcommand{\noopsort}[1]{} \newcommand{\printfirst}[2]{#1}
  \newcommand{\singleletter}[1]{#1} \newcommand{\switchargs}[2]{#2#1}

\appendix

\section{Solution of \eqref{gamPr2}} \label{App0}

We need to prove that \eqref{gamPr2} has a unique solution. The below discussion is an alternative to the solution of this problem given by \cite{EN2001}.

First, consider two extreme cases. If no banks default, then $a_i + \sum_{j\ne i} \gamma_j l_{ji} \ge 1, \ \forall i \in [1,N]$. Obviously, the solution of \eqref{gamPr2} is $\gamma_i = 1, \ \forall i \in [1,N]$.

If all banks default, then \eqref{gamPr2} transforms to the form
\[  F(\boldsymbol{\gamma}) = \boldsymbol{\gamma}, \]
\noindent where $F(\boldsymbol{\gamma})$ denotes the lhs of \eqref{gamPr2}. This is a fixed point problem\footnote{This actually is a linear system of equations. However, we want to solve it using a fixed-point iterations method to later apply this technique to the general \eqref{gamPr2}.} that can be solved by the fixed point iterations method. A sufficient condition for local linear convergence of fixed point iterations is that the Jacobian $J(F(\boldsymbol{\gamma}))$ has to obey the condition 
\begin{equation} \label{condition}
|J(F(\boldsymbol{\gamma}))| < 1.
\end{equation}
To prove this in our case, represent the Jacobian in the explicit form 
\begin{equation*} 
J(F(\boldsymbol{\gamma})) = 
\begin{vmatrix}
0      & l_{21} & l_{31} & \dots & l_{N1} \\
l_{12} & 0      & l_{32} & \dots & l_{N2} \\
\hdotsfor{5} \\
l_{1N} & l_{2N}      & l_{3N} & \dots & 0 \\
\end{vmatrix}
= \left(\prod_{k=1}^N \tilde{L}_{k}\right)^{-1}
\begin{vmatrix}
0      & L_{21} & L_{31} & \dots & L_{N1} \\
L_{12} & 0      & L_{32} & \dots & L_{N2} \\
\hdotsfor{5} \\
L_{1N} & L_{2N}      & L_{3N} & \dots & 0 \\
\end{vmatrix}
\end{equation*}

By definition for any matrix $|M| = ||m_{ij}||, \ i,j \in [1,N]$
\[ det(|M|) = \sum_{\chi \in S_N} \mbox{sgn}(\chi) \prod_{i=1}^N m_{i, \chi_i}, \]
\noindent where the sum is computed over all permutations $\chi$ of the set $S_N = [1,2,...,N]$, see \cite{Bellman}. Since all $L_{ij} \ge 0$ we have
\begin{equation} \label{det}
J(F(\boldsymbol{\gamma})) < \left[\sum_{\chi \in S_N} \prod_{i=1}^N L_{i, \chi_i} \right]
\left[ \prod_{k=1}^N \tilde{L}_{k} \right]^{-1}. 
\end{equation}

Now observe that the numerator in \eqref{det} is a sum of the products of the $N$ elements, and each such a product i) is positive, and ii) has its vis-\`{a}-vis in the denominator. However, the denominator contains also some additional positive terms, for instance $\prod_{i=1}^N L_i$, and therefore, $J(F(\boldsymbol{\gamma})) < 1$. 

As an example, when $N=2$ 
\[ |J(F(\boldsymbol{\gamma}))| = \dfrac{L_{21}}{L_1 + L_{12}} \dfrac{L_{12}}{L_2 + L_{21}} < 1. \]

Thus, we proved that the condition \eqref{condition} is always satisfied. Therefore, by the Banach fixed-point theorem (\cite{FPT2003} the map $F(\boldsymbol{\gamma}) \rightarrow \boldsymbol{\gamma}$ is a contraction mapping on $\boldsymbol{\gamma}$, and this implies the existence and uniqueness of the fixed point since a unit cube where $\boldsymbol{\gamma}$ is defined is a compact metric space.

These two extreme cases naturally give rise to the idea of how to solve \eqref{gamPr2} in general by using a fixed-point iteration method. Given the vector $\boldsymbol{\gamma}$ from the previous iteration, we check the condition $a_i + \sum_{j\ne i} \gamma_j l_{ji} < 1$ for all $i \in [1,N]$. If for some $i=k$ this condition is not satisfied, we put $\gamma_k = 1$ and exclude the equation for $\gamma_k$ from \eqref{gamPr2}. Otherwise, this equation remains in the system. After this step is completed, and for instance, $M$ out of $N$ variables $\gamma$ were set to 1, we solve \eqref{gamPr2} for the remaining $N-M$ variables. The uniqueness of the solution and convergence of the fixed-point iterations follow from the above proof.

\section{Closed form representation of the integral} \label{App1}

Due to various regulators requirements assets of large banks cannot drop below their liabilities, which means that their recovery rates $R$ should be close to 1. In this case when computing, e.g., marginal survival probabilities, the domain $\mathbf{D}_{12}$ in Fig.~\ref{Fig2} becomes a positive quadrant. Finding an analytical solution for survival probability in a positive quadrant with non-zero drift is a long standing problem, that to the best of the authors' knowledge was not solved yet. A relevant literature includes \cite{Lipton2001,Zhou2001, Metzler2010, Patras2008} and references therein.

From a technical prospective we want to compute the integral
\begin{equation} \label{Eq21-1}
Q_{1}(t',X_1^\prime,X_2^\prime) = \int_0^\infty d X_1 \int_0^\infty d X_2 G(\tau',X_1,X_2)
\end{equation}
\noindent where the corresponding 2D Green's function is given by \eqref{Green2D}. The closed form solution for this integral is known when the drift $\xi$ in \eqref{Eq21-1} vanishes, see \cite{Metzler2010} and references therein. However, if $\xi \ne 0$ the closed form solution is not known yet. Here we derive this representation in the form of series of generalized and confluent hypergeometric functions.

First, using polar coordinates $R, \phi$ we rewrite \eqref{Eq21-1} in the form
\begin{equation} \label{trans1}
Q_{1}(t',R^\prime, \phi^\prime) = \dfrac{2}{\varpi \vartheta} e^{\kappa}
\sum_{n=1}^\infty  \sin \left( \nu_{n}\phi ^{\prime }\right)
\int_0^\varpi \sin(\nu_n \phi)d \phi \int_0^\infty R e^{\gamma(\phi) R} e^{- \alpha R^2}
I_{\nu_n}(\beta R) d R
\end{equation}
\noindent where
\begin{equation*}
\kappa = -\frac{\langle \xi^T,\theta \rangle \vartheta}{2} - \frac{R^{\prime 2}}{2\vartheta} - \langle X^\prime, \theta \rangle, \quad \beta = R^{\prime }/\vartheta, \quad \gamma(\phi) = (\theta_2 + \rho \theta_1) \sin \phi + \bar{\rho}\theta_2 \cos \phi.
\quad \alpha = \dfrac{1}{2\vartheta}.
\end{equation*}

Next, we use the Gegenbauer expansion of the complex exponential of two variables in terms of the ultra-spherical (Gegenbauer) polynomials, \cite{Abreu2008}
\begin{equation} \label{Geg}
e^{i x s} = \Gamma(\nu) \left(\dfrac{s}{2}\right)^{-\nu} \sum_{k=0}^\infty i^k (\nu + k) J_{\nu+k}(s) C^\nu_k(x),
\end{equation}
\noindent where $C^\nu_k(x)$ are the Gegenbauer polynomials (\cite{as64}), and the parameter $\nu$ can be arbitrary chosen. It can also be seen as a Neumann series (\cite{Watson1966})  of the exponential $e^{i x t}$. By changing variables $s = i S$ in \eqref{Geg} the latter can be transformed to
\begin{equation} \label{Geg1}
e^{-S x} = \Gamma(\nu) \left(\dfrac{S}{2}\right)^{-\nu} \sum_{k=0}^\infty (-1)^k (\nu + k) I_{\nu+k}(S) C^\nu_k(x).
\end{equation}
Substitution of this representation with $S = \beta R$ and $x = - \gamma(\phi)/\beta$ into \eqref{trans1} yields
\begin{align} \label{trans2}
Q_{1}(t',R^\prime, \phi^\prime) &= \dfrac{2 \bar{\rho}}{\varpi \vartheta} e^{\kappa} \Gamma(\nu)\left(\dfrac{\beta}{2}\right)^{-\nu}
\sum_{n=1}^\infty  \sum_{\mu=0}^\infty (-1)^\mu (\nu + \mu)  \sin \left( \nu_{n}\phi ^{\prime }\right) \\
&\cdot \int_0^\varpi \sin(\nu_n \phi) C^\nu_\mu(-\gamma(\phi)/\beta)d\phi
\int_0^\infty R^{1-\nu} e^{- \alpha R^2} I_{\nu_n}(\beta R) I_{\nu + \mu}(\beta R) d R. \nonumber
\end{align}

For the sake of simplicity,  it does make sense to choose $\nu = 1$, and then use the identity (\cite{GR2007})
\begin{align} \label{iden}
\int_0^\infty & e^{- \alpha R^2} I_{\nu_n}(\beta R) I_{\nu + \mu}(\beta R) d R  =
2^{-\nu_n - \mu -1} \alpha^{-(\nu_n + \mu + 1)/2} \beta^{\nu_n + \mu}
\dfrac{\Gamma[(1 + \mu + \nu_n)/2]}{\Gamma(\mu + 1)\Gamma(\nu_n + 1)} \\
&\cdot \pFq{3}{3}{\dfrac{\nu_n + \mu + 1}{2}, \dfrac{\nu_n + \mu + 2}{2}, \dfrac{\nu_n + \mu + 1}{2}}{\mu + 1,
\nu_n + 1, \mu + \nu_n +1}{-\dfrac{\beta^2}{\alpha}}, \nonumber
\end{align}
\noindent where ${}_3F_3(a_1,a_2,a_3; b_1,b_2,b_3; z)$ is a generalized hypergeometric function (\cite{Askey2010}).

Further, observe that at $\nu = 1$ the Gegenbauer polynomials become the Chebyshev polynomials of the second kind which admit the representation (\cite{as64})
\[
U_n(x) = \sum_{k=0}^{[n/2]} (-1)^k C_k^{n-k} (2 x)^{n-2k},  \]
\noindent where $[x]$ is the floor function. Therefore, the first integral in \eqref{trans2} assumes the form
 \begin{equation} \label{I1}
 I_1 = \sum_{k=0}^{[\mu/2]} 2^{\mu-2k} (-1)^{k} C_k^{\mu-k} \int_0^\varpi \sin(\nu_n \phi) ( \bar{\theta}_1\sin \phi + \bar{\theta}_2 \cos \phi)^{\mu-2k} d \phi,
\end{equation}
\noindent with
\[ \bar\theta_1 = \frac{\theta_2 + \rho \theta_1}{\beta}, \quad \frac{\bar\rho \theta_2}{\beta}, \]
\noindent and $C_k^{\mu-k}$ be the binomial coefficient.

The integral in the rhs of \eqref{I1} can be taken in closed form and reads
\begin{align} \label{I2}
\int_0^\varpi & \sin(\nu_n \phi) ( \bar{\theta}_1\sin \phi + \bar{\theta}_2 \cos \phi)^{\mu-2k} d \phi = \\
& \frac{\omega  2^{2 k-\mu -1}}{\omega ^2 (\mu -2 k)^2-\pi ^2 n^2} \Bigg\{
a_1\left[ b_1 \mathcal{F}_1(n) + b_2 \mathcal{F}_1(-n)\right] +
a_2\left[ b_1 \mathcal{F}_2(n) + b_2 \mathcal{F}_2(-n)\right] \Bigg\} \nonumber \\
\mathcal{F}_1(n) &= {_2F_1}\left(2 k-\mu ,k+\frac{1}{2} \left(\frac{\pi  n}{\omega }-\mu \right),k+\frac{1}{2} \left(\frac{\pi  n}{\omega }-\mu \right) + 1,-1+\frac{2 \bar{\theta}_1}{\bar{\theta}_1-i \bar{\theta}_2}\right), \nonumber \\
\mathcal{F}_2(n) &= {_2F_1}\left(2 k-\mu, k+\frac{1}{2} \left(\frac{\pi  n}{\omega }-\mu \right), k+\frac{1}{2} \left(\frac{\pi  n}{\omega } - \mu\right) + 1, \frac{e^{2 i \omega } (\bar{\theta}_1+i \bar{\theta}_2)}{\bar{\theta}_1-i \bar{\theta}_2}\right), \nonumber \\
a_1 & = -\bar\theta_2^{\mu -2 k} \left(-\frac{i \bar\theta_2}{\bar{\theta}_1-i \bar\theta_2}\right)^{2 k-\mu }, \quad a_2 = e^{-i \pi  n}, \quad b_i = (-1)^{i-1}\omega  (\mu -2 k)+\pi  n, \ i=1,2, \nonumber
\end{align}
\noindent where $_2F_1(a,b,c,x)$ is confluent hypergeometric function (\cite{as64}).

Although in \eqref{I2} the integral is represented as a function of a complex argument, it could be shown that it is real. For example,
\begin{align}
\int_0^\varpi & \sin(\nu_n \phi) ( \bar{\theta}_1\sin \phi + \bar{\theta}_2 \cos \phi) d \phi =  \dfrac{\pi  n \omega}{\omega ^2-\pi ^2 n^2} \left[ (-1)^n (\bar{\theta}_1 \sin (\omega )+ \bar{\theta}_2 \cos (\omega )) - \bar{\theta}_2\right] \nonumber \\
\int_0^\varpi & \sin(\nu_n \phi) ( \bar{\theta}_1\sin \phi + \bar{\theta}_2 \cos \phi)^2d \phi  =  \dfrac{1}{2 \pi ^3 n^3-8 \pi  n \omega ^2} \Bigg\{ -4 \omega ^3(\theta_1^2 + \theta _2^2)  + 2 \pi ^2 \theta _2^2 n^2 \omega  \nonumber \\
&+ (-1)^n \omega  \left[\pi ^2 n^2 \left(\left(\theta _1^2-\theta _2^2\right) \cos (2 \omega )-2 \theta _1 \theta _2 \sin (2 \omega )\right)-\left(\theta _1^2+\theta _2^2\right) \left(\pi ^2 n^2-4 \omega ^2\right)\right] \Bigg\}, \nonumber
\end{align}
\noindent etc.

\section{Computationally efficient representation of \protect\eqref{CDSnew}} \label{App2}

First, let us mention that the problem given in \eqref{cds} is defined in the semi-infinite domain
$X_1 \in [0,\infty), \ X_2 \in [0, \infty)$. However, for practical purposes this infinite domain is always truncated by some reasonably large value $M_i, \ i=1,2$. Thus, we consider \eqref{cds} with $X_2 \in [0, M_2]$. Strictly speaking, this truncation will change the Green's function representation (\cite{Polyanin2002}), however the error should be small when $X_2^\prime \to \infty$, or in other words it is within the truncation error of changing the upper boundary from $\infty$ to $M_2$.

To exactly match the boundary conditions in \eqref{cds}, we replace $C_1(t,X_,X_2)$ with a new function
\begin{equation} \label{newFun}
\tilde{C}_1(t,X_,X_2) = C_1(t,X_,X_2) - \dfrac{X_2}{M_2}c_{1,\infty}(t,X_1) - \left(1 - \dfrac{X_2}{M_2} \right) \Psi(t,X_1)
\end{equation}

Function $\tilde{C}_1(t,X_,X_2)$ solves the following problem:
\begin{align} \label{a2-1}
\tilde{C}_{1,t}(t,X_1,X_2) &+ \mathcal{L} \tilde{C}_1(t,X_1,X_2) = \Xi(t,X_1,X_2),   \\
\tilde{C}_1(t,X_1,0) &= \tilde{C}_1(t,0,X_2) =
\tilde{C}_1(t,X_1,M_2) = \tilde{C}_1(t,\infty,X_2) = 0,  \nonumber \\
\tilde{C}_1(T,X_1,X_2) &= \alpha_1 \mathbf{1}_{(X_1,X_2) \in [\hat{\mathbf{D}} \cup \mathbf{D}_2]}
- \dfrac{X_2}{M}(1-R_1) \mathbf{1}_{X_1 \le \mu_1^=} - \left(1 - \dfrac{X_2}{M} \right) (1-R_1) \mathbf{1}_{X_1 \le \tilde \mu_1^=}. \nonumber
\end{align}

The solution of this problem is given by the formula (\cite{Polyanin2002})
\begin{align} \label{a2_sol}
\tilde{C}_1(t^\prime,X^\prime_1,X^\prime_2) &=
\int_0^\infty d X_1 \int_0^\infty d X_2 \tilde{C}_1(\tau^\prime,X_1,X_2) G(\tau^\prime, X_1,X_2) \\
&+ \int_0^{\tau^\prime} d s \int_0^\infty d X_1 \int_0^\infty d X_2 \Xi(\tau^\prime-s, X_1, X_2)
G(\tau^\prime-s, X_1,X_2). \nonumber
\end{align}

At $X_2 \to 0$ and $X_2 \to \infty$ due to the boundary conditions for $C_1(t,X_1,X_2)$ the new function $\tilde{C}_1(t,X_1,X_2)$ vanishes. Also, according to the boundary conditions $c_{1,\infty}(t,X_1) \to -\varsigma_1 (T-t)$ at $X_1 \to \infty$ as well as  $c_{1,0}(t,X_1)$, and $C_1(t,X_1,X_2)$.  Therefore, in this limit, $\tilde{C}_1(t,X_1,X_2)$ vanishes as well. Finally, at $X_1 = 0$ we have $c_{1,\infty}(t,0) = c_{1,0}(t,0)
= C_1(t,X_1,X_2) = 1-R_1$. Therefore, in this limit, $\tilde{C}_1(t,X_1,X_2) = 0$. Thus, function  $\tilde{C}_1(t,X_1,X_2)$ satisfies the homogeneous boundary conditions.

Now, let us give an exact representation of $\Xi(t,X_1,X_2)$. We need to apply the operator $\partial_t + \mathcal{L}$ to both parts of \eqref{newFun} and take into account \eqref{cds} for $C_1(t,X_1,X_2)$. Omitting a tedious algebra we obtain
\begin{align} \label{source}
\Xi(t,X_1,X_2) &= \sum_{i=1}^4 a_i(t,X_1,X_2) \\
a_1(t,X_1,X_2) &= \delta(X_1 - \tilde{\mu}_1^<) \tilde{a}_1(t, X_2), \quad
\tilde{a}_1(t,X_2) =  \fp{c_{1,0}(t,X_1)}{X_1}\Big|_{X_1 = \tmm} \dfrac{X_2-M_2}{M_2} \nonumber \\
&\equiv d_1(t)X_2 + d_2(t) \nonumber \\
a_2(t,X_1, X_2) &= \delta^\prime(X_1 - \tilde{\mu}_1^<) \tilde{a}_2(t,X_1,X_2), \quad \tilde{a}_2(t,X_1,X_2) = \left[c_{1,0}(t,X_1) - c_{1,0}(t,\tilde{\mu}_1^<)\right]\dfrac{X_2-M_2}{2 M_2}, \nonumber \\
a_3(t,X_1,X_2) &= \mathbf{1}_{X_1 > \tilde{\mu}_1^<} \tilde{a}_3(t,X_1,X_2), \quad
\tilde{a}_3(t,X_1, X_2) = \dfrac{1}{M_2}  \Big[ \xi_2(t) \left(c_{1,0}(t,X_1) - c_{1,0}(t,\tmm) \right) \nonumber \\
&+ (X_2-M_2) \left(\varsigma_1 - \partial_t c_{1,0}(t,\tilde{\mu}_1^<)\right) + \rho \partial_t c_{1,0}(t,X_1) \Big] \equiv X_2 b_1(t,X_1) + b_2(t,X_1),  \nonumber \\
a_4(t,X_1, X_2) &= \tilde{a}_4(t,X_1) = - \varsigma_1 + \dfrac{1}{M_2}
 \Big[ \xi_2(t) \left(c_{1,\infty}(t,X_1) - c_{1,0}(t,\tilde{\mu}_1^<) \right) \nonumber \\
&+  \rho \left(\partial_t c_{1,0}(t,X_1) - \partial_t c_{1,\infty}(t,X_1)\right)\Big].   \nonumber
\end{align}
Further, denote
\[ J_i = \int_0^{\tau^\prime} d s \int_0^\infty d X_1 \int_0^\infty d X_2 a_i(\tau^\prime-s, X_1, X_2) G(\tau^\prime-s, X_1,X_2). \]

By using  \eqref{a2_sol} and \eqref{source} we obtain
\begin{align*}
\int_0^{\tau^\prime} & d s \int_0^\infty d X_1 \int_0^\infty d X_2 \Xi(\tau^\prime-s, X_1, X_2) G(\tau^\prime-s, X_1,X_2) = \sum_{i=1}^4 J_i \\
J_1 &= \int_0^{\tau^\prime} d s \int_0^\infty \tilde{a}_1(\tau^\prime-s, X_1,X_2) G(\tau^\prime-s, \tmm,X_2) d X_2 \\
&= \int_0^{\tau^\prime} d s \left[ d_2(\tau^\prime-s) Y_1(\taups, \tmm) + d_1(\taups) Y_2(\taups, \tmm)\right], \\
J_2 &= \int_0^{\tau^\prime} d s \int_0^\infty  \tilde{a}_1(\taups, \tmm, X_1, X_2) G_{X_1}(\taups, \tmm,X_2) d X_2  \\
&= \int_0^{\tau^\prime} d s \left[ d_2(\taups) Z_1(\taups, \tmm) + d_1(\taups) Z_2(\taups, \tmm)\right], \\
J_3 &= \int_0^{\tau^\prime} d s \int_{\tmm}^\infty d X_1 \int_0^\infty d X_2 \tilde{a}_3(\taups, X_1, X_2) G(\taups, X_1,X_2) \\
&= \int_0^{\tau^\prime} d s \Big[ \int_{\tmm}^\infty d X_1 b_2(\taups,X_1)
Y_1(\taups, X_1)   + \int_{\tmm}^\infty d X_1 b_1(\taups,X_1)
Y_2(\taups, X_1) \Big],
\end{align*}
\begin{align*}
J_4 &= \int_0^{\tau^\prime} d s \int_0^\infty d X_1 \tilde{a}_3(\taups, X_1)
Y_1(\taups, X_1),
\end{align*}
\noindent where
\begin{align*}
Y_1(t, X_1) &= \int_0^\infty G(t,X_1,X_2)d X_2, \qquad
Y_2(t, X_1) = \int_0^\infty X_2 G(t,X_1,X_2)d X_2, \\
Z_1(t, X_1) &= \int_0^\infty G_{X_1}(t,X_1,X_2)d X_2, \qquad
Z_2(t, X_1) = \int_0^\infty X_2 G_{X_1}(t,X_1,X_2)d X_2.
\end{align*}

Also we emphasize that a pretty similar approach can be used for computing marginal probabilities.

\section{Computationally efficient representation of \protect\eqref{FTDnew}} \label{App3}
By using a similar idea as in the previous Appendix we first truncate the infinite domain $(X_1,X_2) \in [0,\infty) \times [0, \infty)$ to a finite domain $(X_1,X_2) \in [0,M_1] \times [0, M_2]$ and introduce a new function
\begin{align} \label{newFunFtd}
\tilde{F}_1(t,X_,X_2) &= F_1(t,X_,X_2) - h(t,X_1,X_2) \\
h(t,X_1,X_2) &= \Bigg\{
\left[\dfrac{X_2}{M_2}f_{1,\infty}(t,X_1) + \left(1 - \dfrac{X_2}{M_2} \right)(1-R_2)\right] \mathbf{1}_{X_1} \nonumber \\
& + (1-R_1)\left(1 - \mathbf{1}_{X_1} \right) \Bigg\}\mathbf{1}_{M_1-X_1}  - f_{2,\infty}(t,X_2)\left[1- \mathbf{1}_{M_1 - X_1}\right]. \nonumber
\end{align}

Function $\tilde{F}_1(t,X_,X_2)$ solves the following problem:
\begin{align} \label{a3-1}
\tilde{F}_{1,t}(t,X_1,X_2) &+ \mathcal{L} \tilde{F}_1(t,X_1,X_2) = \Upsilon(t,X_1,X_2),   \\
\tilde{F}_1(t,X_1,0) &= \tilde{F}_1(t,0,X_2) =
\tilde{F}_1(t,X_1,M_2) = \tilde{F}_1(t,M_1,X_2)= 0,  \nonumber \\
\tilde{F}_1(T,X_1,X_2) &= F_1(T,X_1,X_2) - h(T,X_1,X_2), \nonumber
\end{align}
\noindent and $f_{i,\infty}(T,X_i) = (1-R_i) \mathbf{1}_{X_i \le \mu_i^=}$.

The solution of this problem is given by the formula (\cite{Polyanin2002})
\begin{align} \label{a3_sol}
\tilde{F}_1(t^\prime,X^\prime_1,X^\prime_2) &=
\int_0^\infty d X_1 \int_0^\infty d X_2 \tilde{F}_1(\tau^\prime,X_1,X_2) G(\tau^\prime, X_1,X_2) \\
&+ \int_0^{\tau^\prime} d s \int_0^\infty d X_1 \int_0^\infty d X_2 \Upsilon(\tau^\prime-s, X_1, X_2)
G(\tau^\prime-s, X_1,X_2). \nonumber
\end{align}

In order to compute $\Upsilon(t,X_1,X_2)$ apply the operator $\partial_t + \mathcal{L}$ to both parts of \eqref{newFunFtd} and take into account \eqref{FTD} for $F_1(t,X_1,X_2)$. Omitting a tedious algebra we obtain
\begin{align} \label{U1}
\Upsilon(t,X_1,X_2) &= -2 \varsigma_1 + \frac{1}{2}\left[-1 + R_1 + f_{2,\infty}(t,X_2) \right] \delta'_{X_1}(M_1 - X_1) \\
& - \frac{\delta'_{X_1}(0)}{M_2} \left[ M_2(R_1-R_2) - y(1 - R_2 + f_{1,\infty}(t,X_1))\right] \nonumber \\
&+ \delta(X_1) b_1(t,X_1,X_2) + \delta(M_1 - X_1) b_2(t,X_1,X_2), \nonumber
\end{align}
\noindent where $b_i(t,X_1,X_2)$ are some functions. We omit the explicit form of these functions since the integrals
\begin{align*}
\int_0^\infty \delta(X_1) G(\tau'-s, X_1,X_2) b_1(t,X_1,X_2) d X_1 &= 0,\\
 \int_0^\infty \delta(M_1 - X_1) G(\tau'-s, X_1,X_2) b_2(t,X_1,X_2) &= 0,
 \end{align*}
\noindent due to the boundary conditions for the Green's function. Therefore, the final representation for the boundary integral in \eqref{a3_sol} reads
\begin{align*}
\int_0^{\tau^\prime} & d s \int_0^\infty d X_1 \int_0^\infty d X_2 \Upsilon(\tau^\prime-s, X_1, X_2)
G(\tau^\prime-s, X_1,X_2) = K_1 + K_2 + K_3, \\
K_1 &= -2 \varsigma_1 \int_0^{\tau^\prime} d s \int_0^\infty d X_1 Y_1(\tau'-s,X_1) \\
K_1 &= \dfrac{1}{2} \int_0^{\tau^\prime} d s \int_0^\infty
\left[-f_{2,\infty}(t,X_2) + R_1 - 1\right] G_{X_1}(\tau^\prime-s, M_1,X_2) d X_2 , \\
K_3 &= \int_0^{\tau^\prime} \left[ (R_1 - R_2) Z_1(\tau'-s,0) +
\dfrac{R_1 + R_2 - 2}{M_2} Z_2(\tau'-s,0) \right]d s.
\end{align*}

At numerical (discrete) realization all $K_i, \ i=1-3$ vanish at the boundaries as well as the first integral in \eqref{a3_sol}, and so does $\tilde{F}_1(t,X_1,X-2)$. Therefore, by definition of $\tilde{F}_1(t,X_1,X-2)$ this preserves the correct boundary conditions for $F_1(t,X_1,X-2)$.

\end{document}